\documentstyle[galley,epsf]{mn}
\newif\ifAMStwofonts
\ifoldfss
  \ifCUPmtlplainloaded \else
    \NewTextAlphabet{textbfit} {cmbxti10} {}
    \NewTextAlphabet{textbfss} {cmssbx10} {}
    \NewMathAlphabet{mathbfit} {cmbxti10} {} % for math mode
    \NewMathAlphabet{mathbfss} {cmssbx10} {} %  "   "    "
  \fi
  \ifAMStwofonts
    \ifCUPmtlplainloaded \else
      \NewSymbolFont{upmath} {eurm10}
      \NewSymbolFont{AMSa} {msam10}
      \NewMathSymbol{\upi}     {0}{upmath}{19}
      \NewMathSymbol{\umu}     {0}{upmath}{16}
      \NewMathSymbol{\upartial}{0}{upmath}{40}
      \NewMathSymbol{\leqslant}{3}{AMSa}{36}
      \NewMathSymbol{\geqslant}{3}{AMSa}{3E}

      \let\geq=\geqslant 
    \fi
  \fi
\fi % End of OFSS

\ifnfssone
  \newmathalphabet{\mathit}
  \addtoversion{normal}{\mathit}{cmr}{m}{it}
  \addtoversion{bold}{\mathit}{cmr}{bx}{it}
  \newmathalphabet{\mathbfit} % math mode version of \textbfit{..}
  \addtoversion{normal}{\mathbfit}{cmr}{bx}{it}
  \addtoversion{bold}{\mathbfit}{cmr}{bx}{it}
  \newmathalphabet{\mathbfss} % math mode version of \textbfss{..}
  \addtoversion{normal}{\mathbfss}{cmss}{bx}{n}
  \addtoversion{bold}{\mathbfss}{cmss}{bx}{n}
  \ifAMStwofonts
    \ifCUPmtlplainloaded \else
      \UseAMStwoboldmath
      \makeatletter
      \new@mathgroup\upmath@group
      \define@mathgroup\mv@normal\upmath@group{eur}{m}{n}
      \define@mathgroup\mv@bold\upmath@group{eur}{b}{n}
      \edef\UPM{\hexnumber\upmath@group}
      \new@mathgroup\amsa@group
      \define@mathgroup\mv@normal\amsa@group{msa}{m}{n}
      \define@mathgroup\mv@bold\amsa@group{msa}{m}{n}
      \edef\AMSa{\hexnumber\amsa@group}
      \makeatother
      \mathchardef\upi="0\UPM19
      \mathchardef\umu="0\UPM16
      \mathchardef\upartial="0\UPM40
      \mathchardef\leqslant="3\AMSa36
      \mathchardef\geqslant="3\AMSa3E

      \let\geq=\geqslant 
    \fi
  \fi
\fi % End of NFSS release 1

\ifnfsstwo
  \DeclareMathAlphabet{\mathbfit}{OT1}{cmr}{bx}{it}
  \SetMathAlphabet\mathbfit{bold}{OT1}{cmr}{bx}{it}
  \DeclareMathAlphabet{\mathbfss}{OT1}{cmss}{bx}{n}
  \SetMathAlphabet\mathbfss{bold}{OT1}{cmss}{bx}{n}
  \ifAMStwofonts
    \ifCUPmtlplainloaded \else
      \DeclareSymbolFont{UPM}{U}{eur}{m}{n}
      \SetSymbolFont{UPM}{bold}{U}{eur}{b}{n}
      \DeclareSymbolFont{AMSa}{U}{msa}{m}{n}
      \DeclareMathSymbol{\upi}{0}{UPM}{"19}
      \DeclareMathSymbol{\umu}{0}{UPM}{"16}
      \DeclareMathSymbol{\upartial}{0}{UPM}{"40}
      \DeclareMathSymbol{\leqslant}{3}{AMSa}{"36}
      \DeclareMathSymbol{\geqslant}{3}{AMSa}{"3E}

      \let\geq=\geqslant 
    \fi
  \fi
\fi % End of NFSS release 2

\ifCUPmtlplainloaded \else
  \ifAMStwofonts \else % If no AMS fonts
    \def\upi{\pi}
    \def\umu{\mu}
    \def\upartial{\partial}
  \fi
\fi

\begin{document}
\title[Spectrum of V\,CrA at Maximum and Minimum light]
   {High-resolution spectroscopy of  the   R Coronae Borealis  Star V Coronae Australis\thanks{Based on
observations obtained with (1) The Blanco 4m Telescope at the Cerro Tololo 
Inter-American Observatory, which is operated by AURA,Inc., under contract to the
National Science Foundation of USA, and 
(2) The Harlan J. Smith Telescope of  the W.J. 
McDonald Observatory of the University of Texas at Austin.
}}

\author[N. Kameswara Rao \& David L. Lambert]
       {N. Kameswara Rao$^1$ \& David L. Lambert$^2$\\
       $^1$Indian Institute of Astrophysics, Bangalore 560034, India\\
       $^2$The W.J. McDonald Observatory, The University of Texas, Austin, TX 78712-1083, USA\\
}
\date{Accepted 
      Received ; 
      in original form  }
                                                                                  
\pagerange{\pageref{firstpage}--\pageref{lastpage}}
\pubyear{}

\maketitle

\label{firstpage}

\begin{abstract}

  Optical high-resolution  spectra of the
R Coronae Borealis star V CrA at light maximum and  during minimum light
  are
discussed.
Abundance analysis confirms previous results showing that
V CrA has the composition of the small subclass of R Coronae Borealis (RCB)
stars know as `minority' RCBs, i.e.,  the Si/Fe and S/Fe ratios are 100 times
their solar values.
 A notable novel result for RCBs is the detection of the 1-0 Swan system
 $^{12}$C$^{13}$C
bandhead indicating that $^{13}$C is abundant: spectrum synthesis
shows that
$^{12}$C/$^{13}$C is about  3 to 4.
Absorption line profiles are variable at maximum light with
some lines showing evidence of splitting by about 10 km s$^{-1}$.
A spectrum obtained as the star was recovering from a deep
minimum
shows the presence of  cool C$_2$ molecules with a
rotational temperature
of about  1200K, a temperature suggestive of
gas in which carbon is condensing into soot.
The presence of rapidly outflowing gas is shown by
blue-shifted absorption components
of the Na\,{\sc i}  D and K\,{\sc i} 7698 \AA\ resonance lines.
%Other lines in the same spectrum show evidence
%of infalling gas.

\end{abstract}

\begin{keywords}
Star: individual: R CrB: variables: circumstellar matter :other
\end{keywords}

\section{Introduction}

 R Coronae Borealis stars (here, RCBs) are a rare class of
 peculiar variable stars. The two defining
characteristics of RCBs are (i) a propensity to fade at unpredictable times
 by up to about
8 magnitudes as a result of obscuration by clouds of soot,
 and (ii) a supergiant-like atmosphere that is
very H-deficient and He-rich.
The subject of this paper, V Coronae Australis (V CrA) is
even a peculiar member of this class of peculiar  stars. It is a `minority'
R CrB. The distinction between majority and minority members
was made first by Lambert \& Rao (1994) on the basis of an abundance
analysis of  warm RCBs. The minority RCBs are quite
severely deficient in iron relative to the majority RCBs and to the
Sun but some elements, particularly Si and S, have near-solar
abundances in the minority (and majority) RCBs. This combination
results in some very unusual abundance ratios, for example, the
Si/Fe and S/Fe ratios of minority RCBs are approximately 100 times the solar
ratios. 
V CrA also seems to be an especially  lively producer of dust (Feast et al. 1997).
The realization that V CrA is an unusual RCB
led us to occasional spectroscopic  monitoring of its optical
spectrum.
                                                                                 
 The current  paper discusses
  high-resolution spectroscopic observations obtained 
 on seven occasions between 1989 and 2003  when 
the star was either at or near maximum light or in decline by 3 to 5 magnitudes.
A suitable spectrum at maximum light is subjected here to an
abundance analysis. The previous analysis
(Asplund et al. 2000)
was based on a spectrum obtained during a shallow light minimum.
This spectrum may have been  contaminated  by  phenomena
expected at minimum light (i.e.,
 some lines may have been filled in partially by emission). In
addition, our new spectra cover a broader bandpass at higher resolution and
signal-to-noise ratio than the earlier spectrum.  
Other spectra  show  for the first time for V CrA
 line splitting indicative of the presence of
an atmospheric shock. Spectra taken at minimum light are discussed 
in the context of our detailed
studies of 1995-96 and 2003 minima of R CrB (Rao et al 1999; Rao, Lambert
\& Shetrone 2006).

\begin{figure}
%\begin{minipage}{120mm}
\epsfxsize=8truecm
\epsffile{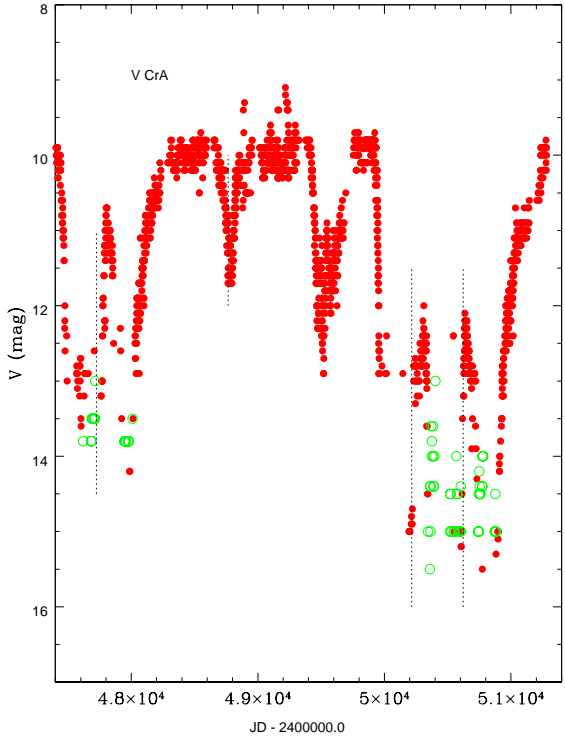}
\caption{The  visual (red dots) light curve of
 V CrA showing the several light  minima during 1988--1998 period.
Dates on which four spectroscopic observations were obtained
are indicated by a dashed line (also see Table 1). Upper limits to the
brightness are shown by green unfilled circles. The observations are from
 the AAVSO database.}
%\end{minipage}
\end{figure}

\section{Observations}

Observations (Table 1) of V CrA were obtained on four occasions with the 
 cross-dispersed echelle spectrograph of the Harlan J. Smith
2.7m reflector at the W.J.  McDonald observatory
(Tull et al. 1995).
 The spectral resolving power, $R = \lambda/d\lambda $,
 employed was 60000. The spectrum covers 3900 to 10000\AA\ with gaps beyond about
5600\AA\ where the echelle orders were incompletely captured on the Tektronix
2048 x 2048 CCD. On three of the four occasions, V CrA was at maximum light
(Figure 2).
   
 Spectroscopic observations (Table 1) were also obtained on three occasions 
 with the Cassegrain
echelle spectrometer on the 4m Blanco reflector at CTIO. The resolving power
 employed was close to 30000 for the 1989 and 1992 spectra
 and about 40000 for the 1996 spectrum. The spectral coverage was  
 5420 \AA\ to 6840 \AA\ during 1989, 5480 \AA\ to 7090 \AA\ in 1992, and
 5750 \AA\ to 8150 \AA\ in 1996. On all the three occasions, the star
 was below its maximum brightness (Figure 1).

\begin{figure}
%\begin{minipage}{120mm}
\epsfxsize=8truecm
\epsffile{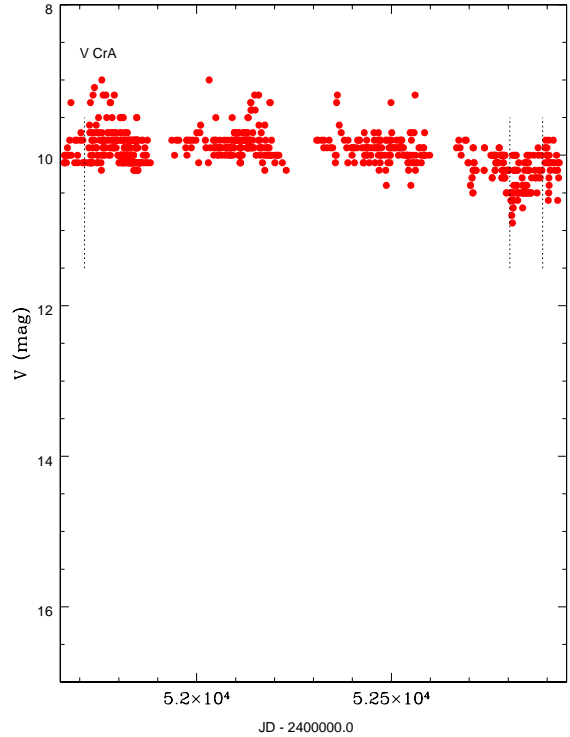}
\caption{The  visual (red dots) light curve of
 V CrA during  2000 - 2003 period.
Dates on which three spectroscopic observations (at maximum light) obtained
are indicated by  dashed lines (also see Table 1). 
 The observations are from
 the AAVSO database.}
%\end{minipage}
\end{figure}

\begin{table*}
\centering
\begin{minipage}{110mm}
\caption{\Large  Spectroscopic Observations of V CrA. }
\begin{tabular}{llrrll}
\hline
                                                                                  
Date & Julian Date & Magnitude & Telescope & Observer$^a$ & Comment\\
(UT) & (2400000+) & V (or vis) \\
\hline
1989 July 16 & 47723.755 & 13 & CTIO\ 4.0m & DLL & recovery\\
1992 May 22 & 48764.838 & 10.9 & CTIO\ 4.0m & DLL\ NKR & decline\\
1996 May 11 & 50214.856 & 14.8 & CTIO\ 4.0m & SB & recovery\\
1997 June 22 & 50621.820 & 13 & McDonald\ 2.7m & DLL\ NKR & recovery\\
2000 June 16 & 51711.779 & 10.0 & McDonald\ 2.7m & DY & maximum\\
2003 June 13 & 52803.876 & 10.3 & McDonald\ 2.7m & BER\ NKR & maximum\\
2003 Sept. 6 & 52888.605 & 10.1 & McDonald\ 2.7m & JS & maximum\\
\hline
\end{tabular}
$^a$ - DLL: D.L.Lambert, NKR: N.K.Rao, SB: S.Balachandran, DY: David Yong,
 BER: B.E.Reddy, JS:Jennifer Simmerer
\label{default}
\end{minipage}
\end{table*}

\section{ V CrA's Spectrum at Maximum Light}
                                                                                 
\subsection{General features}

\begin{figure}
\epsfxsize=8truecm
\epsffile{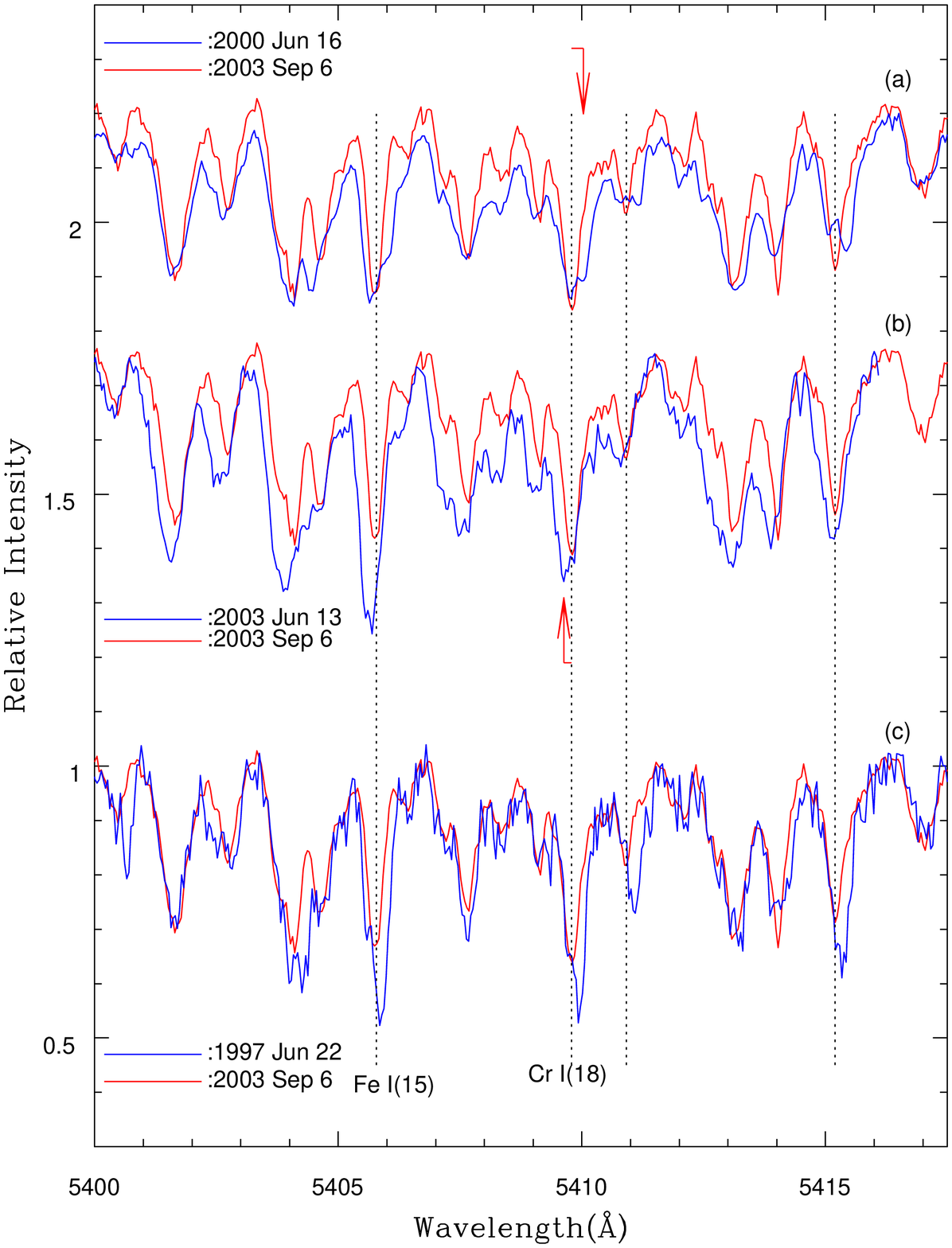}
\caption{Absorption line spectrum of V CrA  near 5410 \AA.
 The spectrum of 2003 September 6 appearing in panels (a), (b), and (c)
 is taken as the normal undisturbed spectrum
at maximum light. This spectrum is compared with  maximum light
  spectra  
(panels (a) and (b)) and a spectrum
obtained at minimum light (panel (c)) that show line doubling.
 Panel (a) shows that,
in the  spectrum obtained on 2000 June
16,  many absorption lines have a weak component  on the redside of
 the principal line. In panel (b),
on the other hand, the spectrum obtained on 2003 June 13 shows the weaker
absorption
 component on the blue side.
 The arrows on top and bottom of panels (a) and (b) highlight
 the red and blue components for the  Cr\,{\sc i} 5409 \AA\ line.
 Panel (c) shows line doubling, particularly in low excitation lines,
in the spectrum obtained on 1997 June 22 during a deep light minimum.}
\end{figure}

Spectra at maximum light
are dominated by
lines of C\,{\sc i}, as expected of an RCB star, and
by lines of  Si\,{\sc i} and S\,{\sc i}  in addition
to the many lines seen in spectra of normal F-G supergiants.
% The Si\,{\sc i} spectrum is very dominant with a majority of
%the lines listed in {\it Revised Multiplet Table} present. 
Even at maximum light, the star shows changes in absorption line profiles,
particularly a variable doubling of some lines, a phenomenon attributable to
pulsations. Percy et al. (2004) found a photometric
 period of 106 days with a range 
 in V of 0.3 magnitudes  during 1987-88. Lawson \& Cottrell (1997) found
 periods ranging from 57 days to 108 days from photometry during 1986--1989.
 However, their radial velocity measurements showed a  variable
 velocity from 14 measurements spanning 13.3 km s$^{-1}$ with
a period of 75 or 125 days. Another notable feature is shown by the
 strongest lines: the profiles of the O\,{\sc i} triplet lines at
 7774 \AA, for example, show extended blue wings indicative of a high
velocity wind.
           The spectra include the C$_2$ Swan system bands.
Unique to V CrA among RCBs observed todate is the clear presence of the 1-0
Swan band at 4744 \AA\ from the $^{12}$C$^{13}$C molecule.

\subsection{Radial velocity and line doubling}

 The 2003
September 6 spectrum shows photospheric absorption lines with symmetrical
profiles. This spectrum is taken as  representative of the undisturbed
stellar atmosphere  and is compared with spectra obtained on other 
occasions. In sharp contrast, many lines in the 2000 June 16 spectrum
appear double
with  the dominant component  accompanied
by a weaker component  to the red (Figure 3). Affected in this way
are lines of Si\,{\sc i}, Ca\,{\sc i},
Sc\,{\sc ii}, Fe\,{\sc i}, and Fe\,{\sc ii}.
In contrast, lines  of C\,{\sc i}, O\,{\sc i}, and S\,{\sc i}, all
of high excitation, do not
show  the red component (Figure 4).
Lines which appear double are increased in equivalent width over the
value when single and symmetric  by no more than about 10\%.
Thus, it appears that the line doubling
is not due to incipient emission emerging in a single stronger and broader line.
Emission lines are not seen in either of  the spectra at maximum light.

\begin{figure}
\epsfxsize=8truecm
\epsffile{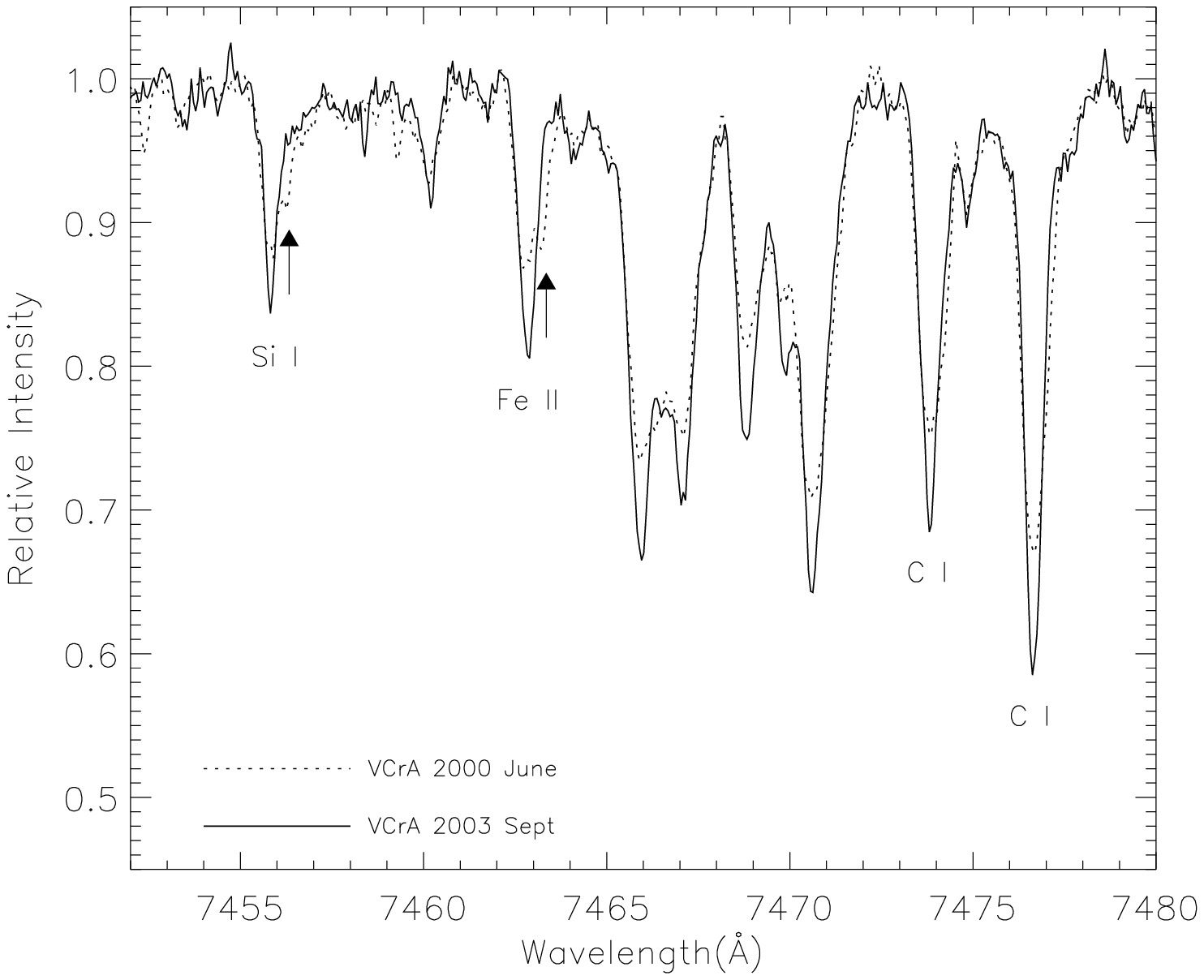}
\caption{Spectrum of V CrA near 7465 \AA\ at maximum light
 on 2003
September 9 (full line) and 2000 June 16 (dotted line).
 Note the presence in the latter spectrum of redshifted
absorption components to the lines of Si\,{\sc i}
and Fe\,{\sc ii}  and their absence for lines of C \,{\sc i}. 
}
\end{figure}

      The spectrum obtained on 2003 June 13, also at maximum light
 shows line doubling, but with  the weaker  component 
shifted to the  blue  of the principal line by about 5 km s$^{-1}$
  (Figure 3 - panel (b)).
 Low excitation lines  
(e.g., Ba\,{\sc ii} 6141 \AA, and Ca\,{\sc i} 6162 \AA)
are much stronger and even
show a  redward extension. 
High-excitation lines remain unaffected except for   a suggestion of
a blueshifted component to C\,{\sc i} and O\,{\sc i} lines.
Line doubling is also seen in the 1997 June 22 spectrum obtained
during a minimum. That the characteristics of the doubling
on 2000 June 16 and 1997 June 22 are similar  suggests that the
photospheric pulsation is present during  a decline. Interruption of the
 pulsation on or about the time of onset of a decline is a possibility
testable only by very intensive spectral coverage.

The mean radial velocity has been reported as
 $-7.9$ km s$^{-1}$ by Lawson \& Cottrell (1997)
and $-8.7$ km s$^{-1}$ by Herbig (1995, private communication). However,
 Skuljan \& Cottrell (2002) while discussing  the V CrA minima of 1994--1998
mention that the  mean radial velocity at light maximum is $-11$ $\pm$2
km s$^{-1}$.  
 The radial velocity is variable,
presumably a reflection of an atmospheric pulsation, with a range of
about 13 km s$^{-1}$ on a timescale of about 50 to 110 days according
to Lawson
\& Cottrell (1997). Herbig's measurements suggest a range of about
$30$ km s$^{-1}$.
Our heliocentric radial velocities of the various components for the
dates of observation  shown in Table 2 are consistent with the published
mean value and the amplitude of the variation. 
 The velocity from
high excitation lines unaffected by line doubling in the 2000 June 16 spectrum
is $-12$ km s$^{-1}$, a value close to the extreme of the previously reported
range. 
 The principal component of the doubled lines in this spectrum
is at this velocity ($-12.6$ km s$^{-1}$). The weaker red component is at
about $+3.6$ km s$^{-1}$ or displaced by about $+11$ km s$^{-1}$ from the
systemic velocity. The spectrum of 2003 June 13 shows the principal
component at a radial velocity of $-4$ km s$^{-1}$ and the blueshifted
components occur at $-18.8$ km s$^{-1}$,  a displacement of
-11 km s$^{-1}$ from the mean.

Line doubling not previously reported for V CrA
is not unknown for RCBs, see, for example,
RY Sgr (Danziger 1963; Cottrell \& Lambert 1982a; Clayton et al. 1994).
How common line doubling is among RCBs is unknown given the paucity of 
high-resolution spectroscopic observations.
Line doubling is 
suggested to be a  sign of shock wave propagation. Since the sound speed in
the atmospheres of these RCBs of $T_{\rm eff}$ 6500K is less than 
10 km s$^{-1}$, amplitudes exceeding these velocities are expected to
produce shocks.
% which might lead to  dust 
%nucleation (Woitke et al. 1997).
 Based on Herbig's (1995) radial 
velocity measurements of V CrA
which showed a range of 30 km s$^{-1}$, Lawson et al (1991) anticipated 
that  V CrA would show evidence for shock waves. Although line doubling
showing inflow and outflow is present with a range in velocity of at least 22 
km s$^{-1}$, more high resolution spectroscopic
observations are needed to trace the shock through the pulsation
cycle. If we assume the average
photometric period of 75 days is valid, the epochs of blue and redshifted
components in Table 2 are separated by about half the period.

\begin{table*}
\centering
\begin{minipage}{180mm}
\caption{\large Radial Velocities (km s$^{-1}$) of various features in V CrA spectrum at various times.}
\begin{tabular}{lrrrrrrr}
\hline
\multicolumn{1}{c}{Feature}&\multicolumn{1}{c}{}&\multicolumn{1}{c}{ }&\multicolumn{1}{c}{Date}&
\multicolumn{1}{c}{} &\multicolumn{1}{c}{} &\multicolumn{1}{c}{} &\multicolumn{1}{c}{}\\
\cline{2-8} \\
&\multicolumn{1}{c}{1989 Jul 16 }&\multicolumn{1}{c}{1992 May 22 }& \multicolumn{1}{c}{1996 May 11 }
& \multicolumn{1}{c}{1997 Jun 22 }&\multicolumn{1}{c}{2000 Jun 16}&\multicolumn{1}{c}{2003 Jun 13}&\multicolumn{1}{c}{2003 Sept 6}\\
\hline
           & min   &  min    & min   &  min   &  max & max & max \\
           &       &         &       &        &      &     &      \\
{\bf Absorption lines}:   &       &         &       &        &      &     &      \\
 Stellar lines& $-$11.6(35) & $-$11.0(24)$^c$& $-$10.3(33)& $-$5.6(18)& $-$12.6(42)&
 $-$4.1(18)& $-$7.5 (25)\\
           & $\pm$2.2& $\pm$2.2& $\pm$1.5 & $\pm$0.5 & $\pm$1.7 &$\pm$1.1 &$\pm$1.2 \\
  Shifted Comp.&     &         &          & 4.0 (27) & 3.6 (22)& $-$18.8(31)  &       \\
               &     &         &          & $\pm$1.3 & $\pm$0.5 & $\pm$2.3        &       \\
 Phillip C$_2$ (3,0)    &      &      &        & $-$10.2(8)   &    & &        \\
           &            &             &           & $\pm$1.8  &    & &        \\
{\bf Sharp emissions}:&        &      &          &          &          &         &       \\
           & $-$4.9 (5) &      &          & 4.4 ( 2)&           &         &     \\
           &            &      &           &             &     &          &     \\
           &            &             &           &           &     & &       \\
{\bf Absorption Components}:&  &      &      &           &                 & & \\
 (Shell)   &            &             &           &           &          & &     \\
 Na\,{\sc i} D  &$-$123 & $-$277      & $-$213    & $-$220$^b$&          & &    \\
           &            &             & $-$183    & $-$190    &           & &   \\
           &            &             &           & $-$160    &          & &    \\
           &            &             &           & $-$154    &          & &    \\
           &            &             &           &           &    & &        \\
 K\,{\sc i}&            &             & $-$214    & $-$220    &          & &    \\
           &            &             &           & $-$194    &     & &         \\
{\bf Broad Na\,{\sc i} emission} & $-$85 to 103$^a$ &   & $-$123 to 103 &  &      &    & \\
           & &             &           &  &       & &     \\
\hline
\end{tabular}
$^a$ The emission is affected by shell absorption on the blue side.\\
$^b$ The absorption extends from $-$240 to $-$130 km s$^{-1}$.
\\
$^c$The number given in parentheses refers to the number of lines used\\
\end{minipage}
\end{table*}

\subsection{Abundance Analysis}

In the earlier comprehensive abundance analysis of RCBs (Asplund et al.
2000), the analysis for V CrA was based on a spectrum of lower resolution
and signal-to-noise than  those available here. Therefore, we undertook
an abundance analysis using the maximum light spectrum of 2003 September 6.
%Equivalent widths of  unblended lines were measured in the selected
%spectra.
 The maximum light spectra of 2000 June 16 and 2003 June 13 were not
considered because many lines are doubled in these spectra.

    The abundance analysis of V CrA followed procedures developed by
Asplund et al. (2000). The analysis used
 the line-blanketed hydrogen-deficient
 model atmospheres described by Asplund et al. (1997a).
An abundance ratio C/He = 1\% by number
of atoms was assumed.
Ionization equilibrium was demanded using   Fe\,{\sc i}/Fe\,{\sc ii},
Mg\,{\sc i}/Mg\,{\sc ii}, Ca\,{\sc i}/Ca\,{\sc ii} and Si\,{\sc i}/Si\,{\sc ii}
to  provide loci in the $T_{\rm eff}$ - $\log g$ plane.
 Excitation equilibrium  through the use of  [O\,{\sc i}] and
high-excitation O\,{\sc i} lines  provides a gravity-insensitive
temperature indicator. We also used the recipe
suggested by Asplund et al. (2000) which adopts a  $M_{\rm bol}$ of R CrBs
  to
 obtain a relation between $T_{\rm eff}$ and log $g$. The
adopted parameters    are indicated   in Figure 5 along with the loci of various
 indicators in the $T_{\rm eff}$ and $\log g$ plane.

\begin{figure}
\epsfxsize=8truecm
\epsffile{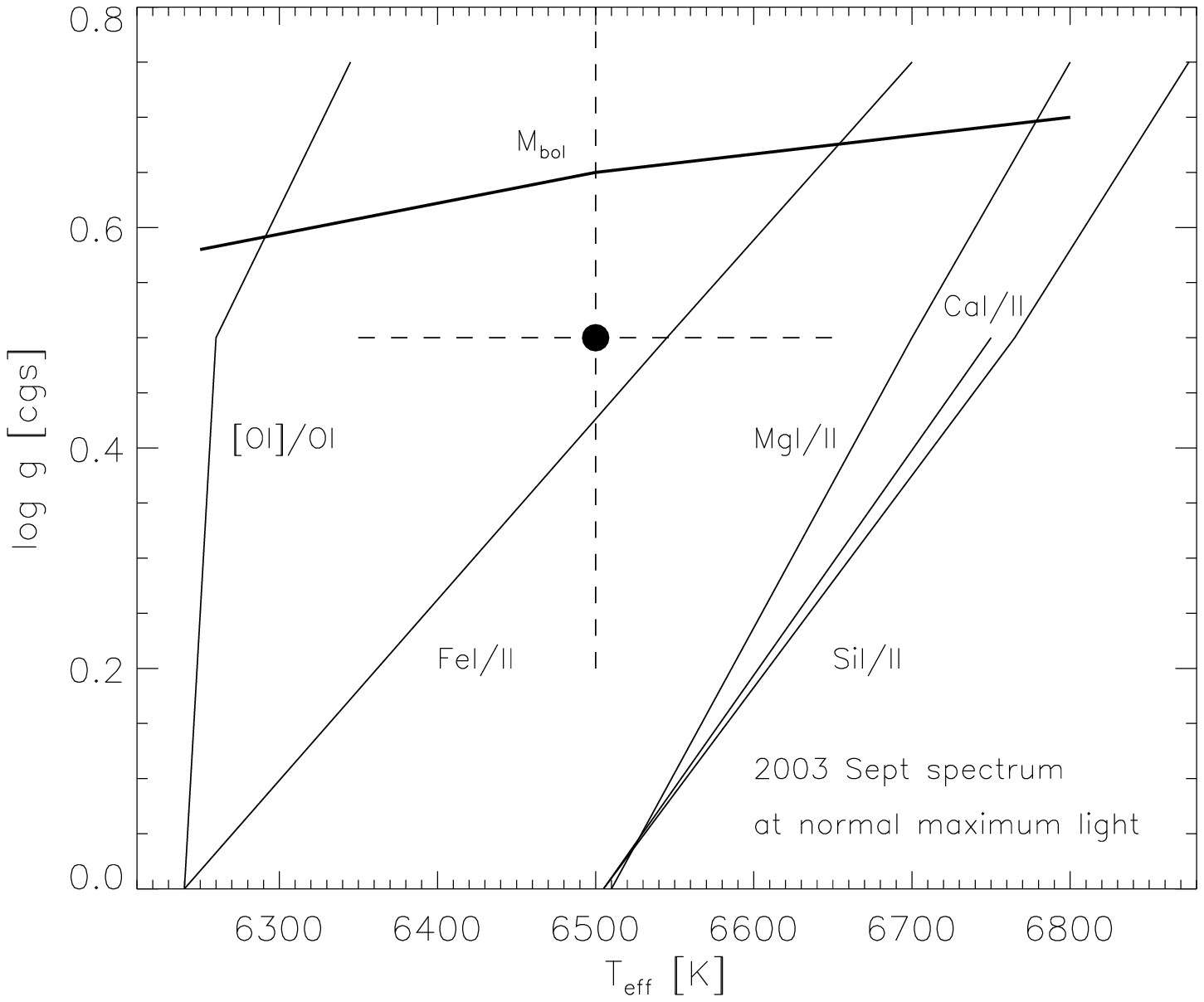}
\caption{Loci of ionization equilibria and other parameters in $T_{\rm eff}$
and $\log g$ plane from the spectrum on 2003 September 6. The final choices of
 $T_{\rm eff}$ and $\log g$ are indicated by the large dot.}
\end{figure}

  The stellar parameters 
$T_{\rm eff}$=6500$\pm$150 K, $\log g$=0.5$\pm$0.3 (cgs units)
were chosen from Figure 5. A
 microturbulence $\xi$$_{\rm tur}$ = 7.0$\pm$1.0 km s$^{-1}$ was found by
the customary requirment that lines of a given species provide an abundance
that is independent of equivalent width.
   The derived abundances are given in Table 3 along with the solar abundances
from Lodders (2003). Differential abundances with respect to the Sun
are given as [El] and [El/Fe]. The sensitivity of the
abundances to a change of $T_{\rm eff}$ by 250 K and $\log g$ by 0.5 dex
is also given.
The abundances are in good agreement with those  listed
by Asplund et al. (2000) except for three elements H, N and O. The present
analysis utilises more  unblended lines than were used earlier and also
provides an upper limit to the abundance of lithium.

\begin{table*}
\centering
\begin{minipage}{120mm}
\caption{ \Large
  Elemental Abundances for V Cr A}

\small\begin{tabular}{lcllcrll}
\hline\hline
 &Log $\epsilon$$_{\rm *}$  &  & \underline{$\delta$ $T_{\rm eff}$, $\delta$ $\log g$ } & n $^{a}$ &
Sun $^{b}$ & [El]$^{c}$  & [El/Fe]
\\
\cline{2-3} Species  & Asplund et al. & present & 250 K, 0.5 & \\
\hline
H\,{\sc i} & 8.0 & 8.68 & 0.17, 0.09 &  1 & 12.00 & -3.3 & -1.3 \\
Li\,{\sc i} &  &   \hspace{-.4cm} $<$0.9  &  &  1 & 3.35 \\
C\,{\sc i} & 8.6 & 8.79 $\pm$ 0.37 & 0.01, 0.01 & 13 & 8.46 & 0.3 & 2.3 \\
N\,{\sc i} & 8.6 & 8.03 $\pm$ 0.54 & 0.15, 0.10 &  5 & 7.90 & 0.1 & 2.1 \\
O\,{\sc i} & 8.7 & 7.78 $\pm$ 0.22 & 0.11, 0.06 &  5 & 8.76 & -1.0 & 1.0 \\
Na\,{\sc i} & 5.9 & 5.67 $\pm$ 0.03 & 0.14, 0.10 & 2 & 6.37 & -0.7 & 1.3 \\
Mg\,{\sc i} & 6.6 & 6.60 $\pm$ 0.28 & 0.14, 0.09 & 4 & 7.62 & -1.0 & 1.0 \\
Mg\,{\sc ii} &     & 6.81           & 0.08, 0.06 & 1 &      & -0.8 & 1.2 \\
Al\,{\sc i} & 5.3 & 5.35 $\pm$ 0.22 & 0.13, 0.12 & 2 & 6.54 & -1.2 & 0.8 \\
Si\,{\sc i} & 7.6 & 7.54 $\pm$ 0.40 & 0.15, 0.10 & 12 & 7.61& -0.1 & 1.9 \\
Si\,{\sc ii} &    & 7.85 $\pm$ 0.30 & 0.12, 0.10 & 3 &     & 0.2 & 2.2  \\
S\,{\sc i} & 7.5 & 7.23 $\pm$ 0.12  & 0.04, 0.08 & 7 & 7.26 & 0.0 & 2.0 \\
K\,{\sc i} &     & 4.81             & 0.22, 0.16 & 1 & 5.18 & -0.4 & 1.6 \\
Ca\,{\sc i} & 5.1 & 5.22 $\pm$ 0.30 & 0.20, 0.09 & 7 & 6.41 & -1.2 & 0.8 \\
Ca\,{\sc ii} &     & 5.48           & 0.02, 0.06 & 1 &     & -0.9 & 1.1 \\
Sc\,{\sc ii} & 2.8 & 2.98 $\pm$ 0.12 & 0.11, 0.05 & 3 & 3.15 & -0.2 & 1.8 \\
Ti\,{\sc ii} & 3.3 & 3.30 $\pm$ 0.17 & 0.07, 0.10 & 3 & 5.07 & -1.7 & 0.3 \\
Fe\,{\sc i} & 5.5 & 5.50 $\pm$ 0.38 & 0.20, 0.10 & 13 & 7.54 & -2.0 & 0.0 \\
Fe\,{\sc ii} &    & 5.54 $\pm$ 0.41 & 0.02, 0.15 &  8 &    & -2.0 & 0.0  \\
Ni\,{\sc i} & 5.6 & 4.87 $\pm$ 0.32 & 0.14, 0.06 &  4 & 6.29 & -1.4 & 0.6 \\
Zn\,{\sc i} & 2.9 & 3.9          &                 &  1 & 4.70 & -0.8 &1.2 \\
Y\,{\sc ii} & 0.6 & 1.06 $\pm$ 0.19 & 0.08, 0.08 & 4 & 2.28 & -1.2 & 0.8 \\
Zr\,{\sc ii} &    & 1.43 $\pm$ 0.37 & 0.06, 0.10 & 3 & 2.67 & -1.2 & 0.8 \\
Ba\,{\sc ii} & 0.7 & 0.30 $\pm$ 0.28 & 0.22, 0.04 & 3 & 2.25 & -2.0 & 0.0 \\
\hline\hline
\end{tabular}
$^{a}$ n = number of lines used in the analyses

$^{b}$\ Recommended abundances for the solar system from Lodders (2003,  Table 2).

$^{c}$\ [El] = log $\epsilon$$_{\rm *}$ - log $\epsilon$$_{\odot}$

\end{minipage}
\end{table*}

Our abundance analysis fully confirms Lambert \& Rao's (1994) identification
of V CrA as a member of their `minority' class for RCBs. Indeed, the
identification is evident from direct inspection of the
spectrum: i.e., Si\,{\sc i} lines are numerous and
  strong and Fe\,{\sc i} and Fe\,{\sc ii}
lines are weak relative to their strengths in a majority RCB of 
comparable temperature. A  mark of a
minority RCB is a low Fe abundance but a near-solar abundance of Si and
S: Table 3 shows that [Si/Fe] $\simeq$ [S/Fe] $\simeq 2$, both  remarkable
ratios.

\subsection{The $^{12}$C/$^{13}$C ratio}

The $^{12}$C$_{\rm 2}$ Swan  bands are fairly strong in the
spectrum of V CrA at maximum light; bandheads of the 0-0 band at 5165 \AA, the 1-0 band at 4737 \AA, and the 0-1 band at 6174 \AA\ are readily
identified.
An absorption feature near 4745 \AA\
(Figure 5) coincides with the
 $^{13}$C$^{12}$C
 1-0 bandhead.
 The 4747 \AA\ feature is similarly strong in the
spectrum of Sakurai's object (V4334 Sgr) for which a low $^{12}$C/$^{13}$C
ratio has been demonstrated (Asplund et al. 1997b).
 This feature is, however,  not
present in spectra of either V854 Cen  or R CrB,
RCB stars with C$_{\rm 2}$ bands of comparable
strength to those in V CrA.

\begin{figure}
\epsfxsize=8truecm
\epsffile{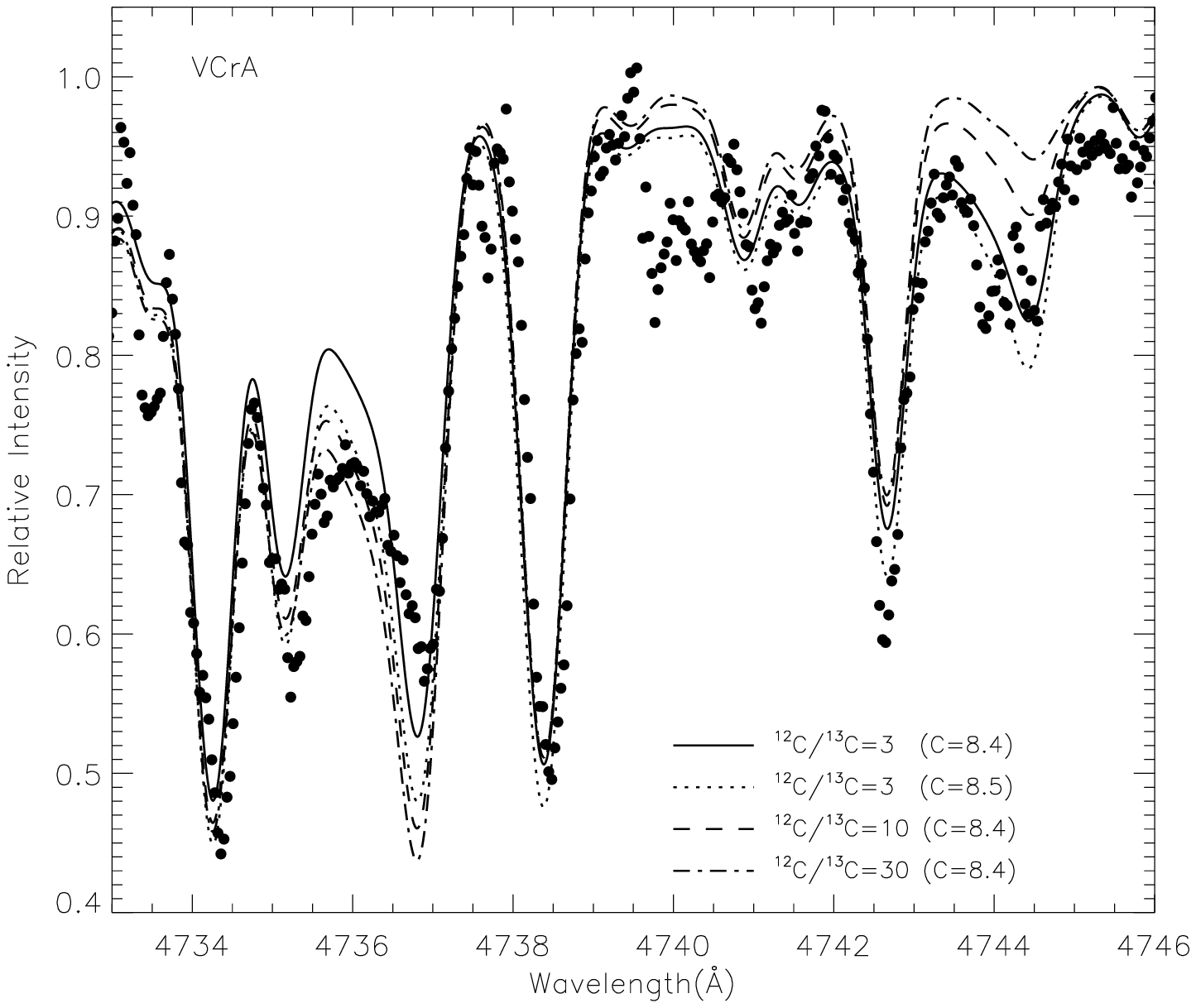}
\caption{The spectrum of V CrA in the region of
 the $^{12}$C$^{12}$C (4736.5 \AA)
and $^{12}$C$^{13}$C (4744.7 \AA) 1-0 Swan bandheads. The observed spectrum
 on 2003 September
6 is shown as
dots. The lines show the synthetic spectrum for various ratios
 of the $^{12}$C/$^{13}$C ratio.  
 }
\end{figure}

%\begin{figure}
%\epsfxsize=8truecm
%\epsffile{v854cenc13c4744A.ps}
%\caption{The spectrum of V854 Cen in the region of the
% $^{12}$C$^{12}$C (4736.5 \AA) and $^{12}$C$^{13}$C (4744.7 \AA)
%1-0 Swan bandheads.
% The observed spectrum at maximum light is shown as dots. Synthetic
% spectra for various $^{12}$C/$^{13}$C ratios
%suggest a lower limit of 30 for V854 Cen. }
%\end{figure}

\begin{figure}
\epsfxsize=8truecm
\epsffile{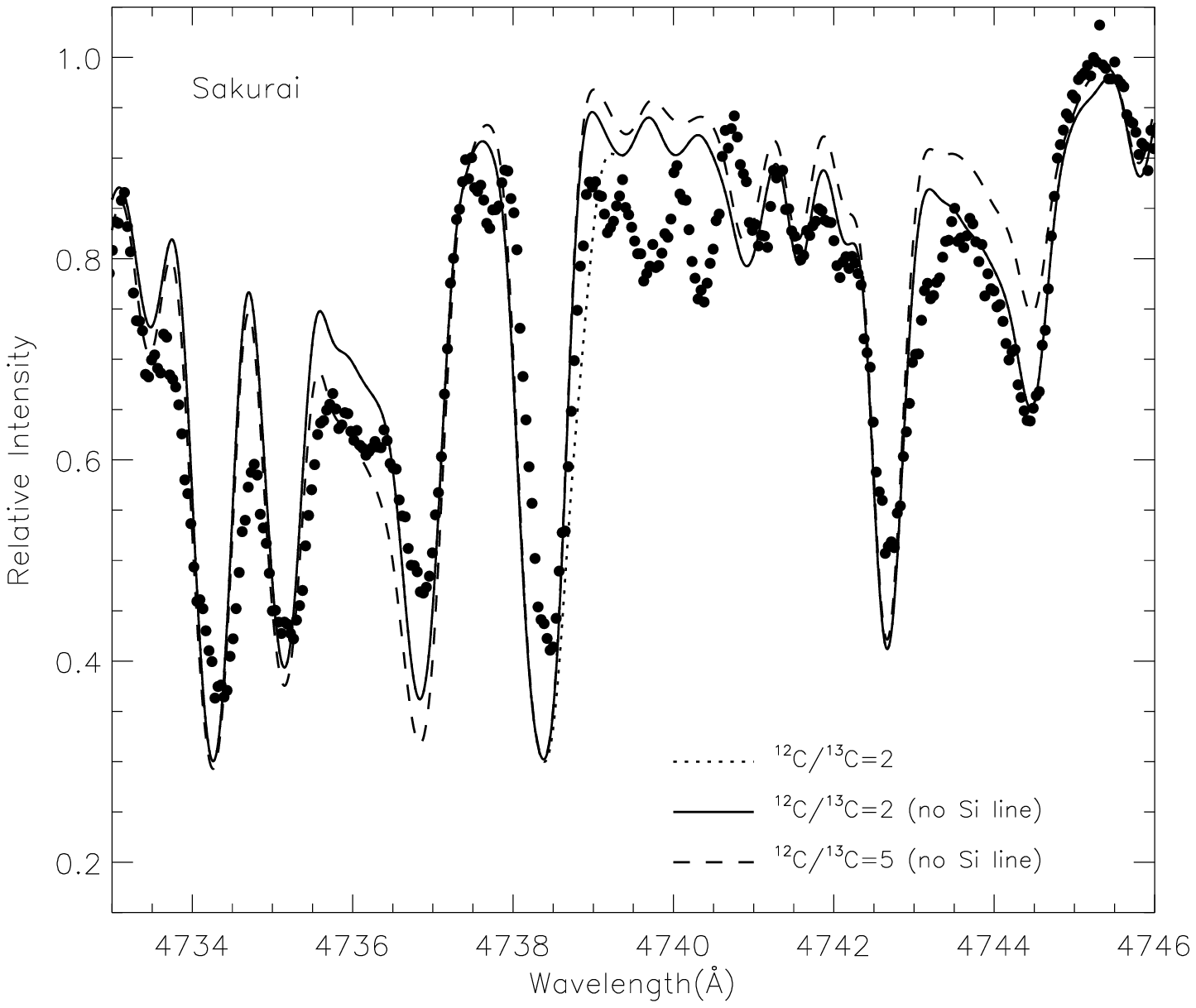}
\caption{The spectrum of Sakurai's object on 1997 October 7 (dots) in the region
of the $^{12}$C$^{12}$C and $^{12}$C$^{13}$C 1-0 Swan bandheads
and synthetic spectra for various $^{12}$C/$^{13}$C ratios (Asplund et al. 1997b).
 }
\end{figure}

We synthesized the 4732--4746 \AA\ interval
that includes the $^{13}$C$^{12}$C and the $^{12}$C$^{12}$C 1-0 bands.
 The line list
 kindly supplied by Martin Asplund  was used previously
 to estimate the $^{12}$C/$^{13}$C
 ratio for Sakurai's object.
 The $^{12}$C$^{13}$C
lines of the 0-0 and 0-1 bands and others in these sequences are present but
intermingled with $^{12}$C$^{12}$C lines. The vibrational isotopic
wavelength shift places the 1-0 $^{12}$C$^{13}$C bandhead to the
red of the blue-degraded  $^{12}$C$^{12}$C and so provides a
fine 8 \AA\ interval for synthesizing $^{12}$C$^{13}$C lines free of blending with $^{12}$C$^{12}$C lines.

Spectrum syntheses for V CrA are shown in Figure 6 for the 2003
September 6 maximum light spectrum. These use the model atmosphere and
abundances listed in Table 3 and different $^{12}$C/$^{13}$C ratios.
 The  C\,{\sc i} and
 molecular C$_{\rm 2}$ lines are  fitted with the same C abundance.
A fit was also made to the
2000 June 16 spectrum in which low excitation
atomic lines exhibit line doubling  but the molecular
lines appear not to be doubled.
 Both spectra  suggest a $^{12}$C/$^{13}$C ratio of 3 to 4.

%\begin{figure}
%\epsfxsize=8truecm
%\epsffile{vcra2kc13c4744Asi.ps}
%\caption{Same as Fig. . The spectrum of V CrA obtained on 2000 June 16 is
%matched to the computed spectrum. }
%\end{figure}

By way of a check, we synthesised  the 4732--4746 \AA\ interval
to fit McDonald spectra of Sakurai's object rich in $^{13}$C.
 Synthesis for Sakurai's object confirms that
the $^{12}$C/$^{13}$C ratio is low: syntheses with ratios of 2 and 5
are  shown in Figure 7.\footnote{Spectrum
 synthesis was undertaken  for V854 Cen.
 A model with C/He of 1\% and
the  abundances listed by Asplund et al. (1998)
provides a  fit to  both the C$_2$ and C\,{\sc i}
 lines. Syntheses suggest a lower
limit of $^{12}$C/$^{13}$C of
 30 for V854 Cen (Rao 2005).}

\subsection{The stellar wind}

        Clayton et al. (2003) reported  a weak P-Cygni profile for 
the He\,{\sc i} 10830 \AA\ line at maximum light in V CrA suggesting 
a shell of hot gas moving outward  with a 
velocity of 295 km s$^{-1}$. The source of
 excitation
for the hot gas and its connection to the photosphere of the star are not clear.
Rao, Lambert \& Shetrone (2006) showed that in R CrB the strong
 photospheric lines,
particularly the  O \,{\sc i} 7771 \AA\ line, had a pronounced blue
wing suggesting mass loss with an expansion velocity of 120 km s$^{-1}$.
The O\,{\sc i} line provides the link between the hot gas responsible
for the  He\,{\sc i} line and the stellar photosphere.
V CrA  and R CrB show similar 
profiles of the O \,{\sc i} 7771 \AA\ line. Figure 8 illustrate the profiles on
three occasions at maximum light and two occasions at minimum light.

\begin{figure}
\epsfxsize=8truecm
\epsffile{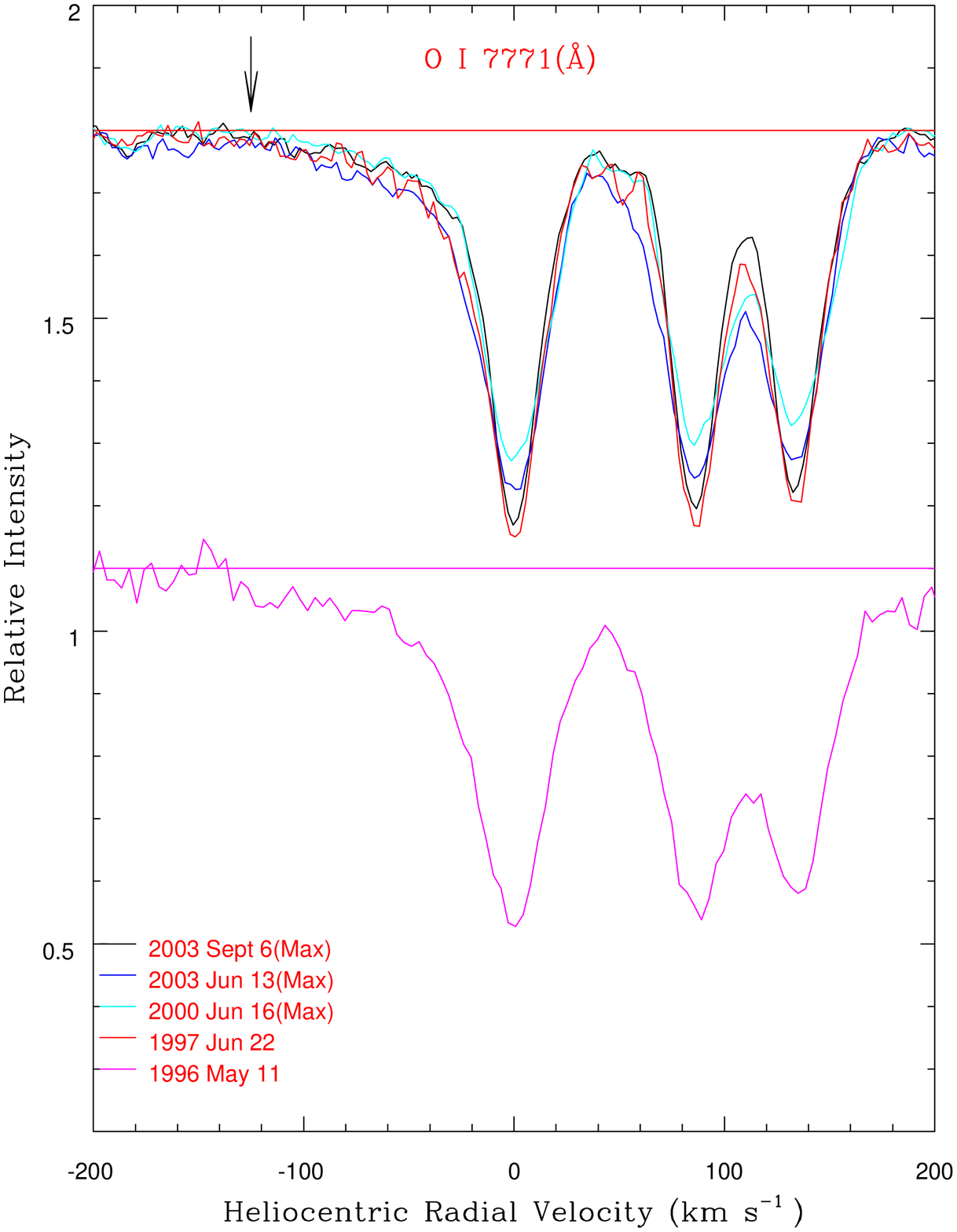}
\caption{Profiles of the O \,{\sc i} triplet lines at 7774\AA. Zero radial
velocity  here corresponds to the stellar velocity of the 7771\AA\ line.
 The top panel
illustrates the profiles obtained from McDonald spectra on three occasions at
maximum light and at minimum on 1997 June 22. The bottom panel illustrates
the  same lines obtained at a lower resolution at CTIO during the
light minimum in 1996 May. Note the pronounced blue wing extending to the
a radial velocity of -125 km s$^{-1}$ relative to the star (shown by the arrow)
on all occasions.}
\end{figure}

The stellar wind for RCBs including V CrA will be discussed
in a future paper but it is significant to note that, since the 
stellar wind profiles are detected on all relevant spectra of V CrA, the
wind is possibly a permanent characteristic.
We do not have 
observations of V CrA during the early decline  from maximum light.
Observations at the 1995--1996 minimum of R CrB suggest that the wings of the
O \,{\sc i} lines are undisturbed even though the core is affected by 
transient emission in the early decline (Rao et al. 1999)

\begin{figure}
\epsfxsize=8truecm
\epsffile{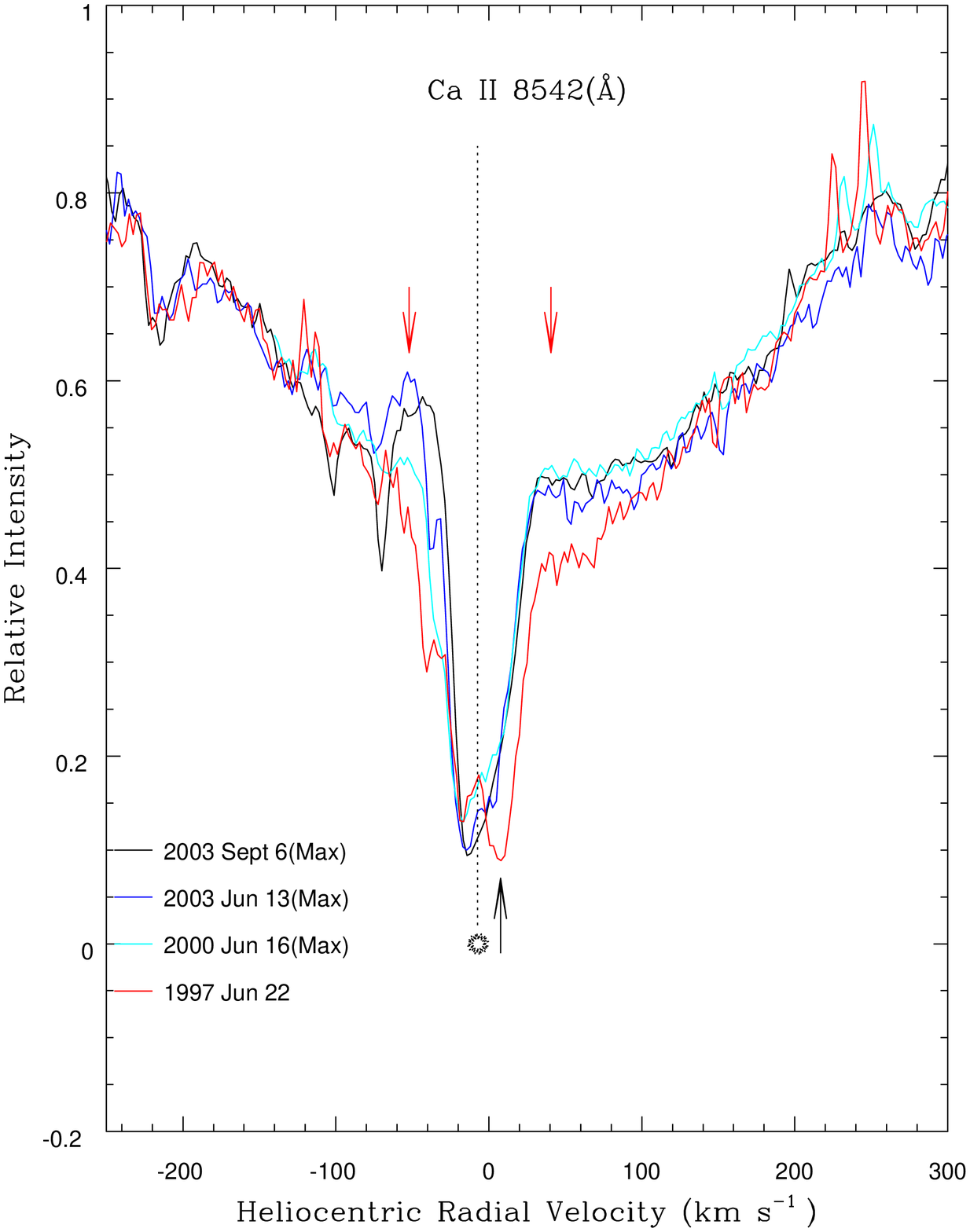}
\caption{The profiles of Ca\,{\sc ii} 8542\AA\ line on three occasions at maximum
light and during the minimum in 1997 June (red line). All three maximum profiles
look similar. The centre of the line shows an asymmetrical absorption core
flanked by   emission bumps on either side (marked by red arrows pointing down).
The profiles are aligned to the stellar absorption lines of the
2003 September 6 spectrum;
the vertical dashed line refers to the stellar velocity. The minimum profile
does not show the emission bumps instead it shows an extra absorption component
(marked by an arrow pointing upwards)
redward of the stellar radial velocity 
(the blue side of the profiles has terrestrial water vapour lines superposed).} 
\end{figure}

       Another permanent feature in the maximum light spectrum of V CrA seems
to be emission cores in the Ca\,{\sc ii} triplet lines.
 The 8498 \AA\ line is in the gap between orders and the 8662\AA\ line
is affected by terrestrial OH lines. Figure 9 illustrates the profiles of the
8542 \AA\ line during three occasions at maximum light and at minimum light in
1997 June. 
     The profiles of the line at maximum light appear to be similar to  an 
absorption core at stellar velocity flanked by emission peaks approximately
symetrically placed at -52 and $+$40 km s$^{-1}$ relative to the stellar
velocity of -7.5 km s$^{-1}$. The emission strength might be variable slightly
but appears to be a permanent feature and unaffected by the appearance of
line doubling.
 The only profile obtained at minimum light shows  no
emission peaks but the presence of a 14.5 km s$^{-1}$ redshifted 
absorption component. 
 The profile is reminiscent of   chromospheric emission and 
 very similar to the  profiles seen in R CrB at light maximum (Rao,
Lambert \& Shetrone 2006). Perhaps, this emission comes from the base of the
stellar wind.

\section{Spectrum at Minimum Light}

For R CrB and other well observed RCBs in decline, emission
lines dominate the optical spectrum.
Two broad classes of emission lines are present:
 a rich set of sharp lines (FWHM $\sim 12$
km s$^{-1}$), and a sparse and diverse set of broad lines (FWHM $\sim 300$ km s$^{-1}$)
(Herbig 1949; Payne-Gaposchkin 1963; Alexander et al. 1972; Rao et al.
1999).
Sharp lines, primarily  low-excitation transitions
of singly-ionized  and neutral metals,
appear very early in the decline and disappear late in the
recovery to maximum light (Rao et al. 1999).
 The broad lines, which are seen only when a RCB has
faded by several magnitudes,
 may include lines of the He\,{\sc i} triplet series, Ca\,{\sc ii} H and K,
 K\,{\sc i} resonance lines at 7664 \AA\ and 7699 \AA,
Na\,{\sc i} D lines, [O\,{\sc ii}], and [N\,{\sc ii}] lines, i.e., a mix of high and
low excitation lines with similar but differing profiles (Rao et al. 1999).
The range of excitation/ionization of these carriers and the differing velocity
and width of the
profiles  suggests that there may be perhaps three regions and/or
excitation mechanisms responsible for the broad
emission lines.

The photospheric absorption line spectrum may also change during a decline.
 In deep minima, the photospheric absorption
lines may be `veiled', i.e., the lines become
very shallow and broad.
New absorption features may also appear.
 Broad blueshifted absorption (`shell') components 
have been seen to accompany commonly  the Na  D lines, and  occasionally
the  K\,{\sc i} 7664 \AA\ and 7699 \AA\ resonance lines,
and the Ca\,{\sc ii} H and K lines.
The Na D absorption components appear especially at and following minimum
light.
The ubiquity of the  sharp and broad emission and the blue-shifted absorption Na D lines
across the sample of RCBs is unknown at present. Probably,
the sharp emission lines of ionized and neutral metals are a common
feature of all declines of all RCBs (Skuljan \& Cottrell 2004).\footnote{Emission
lines of
C\,{\sc i} and O\,{\sc i} first seen early in R CrB's 1995 decline
may have been missed in other declines of R CrB and other RCBs simply
for lack of appropriate observations at early times.}
Since few  RCBs have  been observed in deep minima,  reported
sightings of  broad lines are rare.
 One observation worthy of particular
note, if applicable to all RCBs, is Whitney et al.'s (1992)
 discovery that the Na D broad emission from V854 Cen is unpolarized at a time when the continuum is
markedly polarized. This  suggests that during a decline the stellar 
continuum is seen only in
the reflected light where as the Na D broad emission region is
 viewed directly.

\subsection{Spectra of V CrA at minimum light}

            Our collection of high resolution spectra sample
 four light minima over 8 years. We obtained one spectrum in each
 minimum mostly at the begining of light recovery phase. We  present a 
 description of these spectra.

  {\bf 1989 July 16} : The spectrum was obtained 
after about 288 days since the begining of the minimum. The star reached
a minimum magnitude of 13.8 and was observed near minimum. The spectrum is
 mostly in absorption with few sharp emissions seen in the centres of deep
absorption lines.
 Lines of Sc\,{\sc ii},  and Ba\,{\sc ii} are 
in emission. Only one Sc\,{\sc ii} 6604 \AA\ line is seen above the continuum
in the spectral region covered (Figure 10). The sharp emission lines are all of low 
excitation.
                An unusual aspect of the sharp emissions in this spectrum is
 that they are redshifted relative to the absorption lines (Table 2); 
a blueshift of a few km s$^{-1}$ is common among other RCBs.

\begin{figure}
\epsfxsize=8truecm
\epsffile{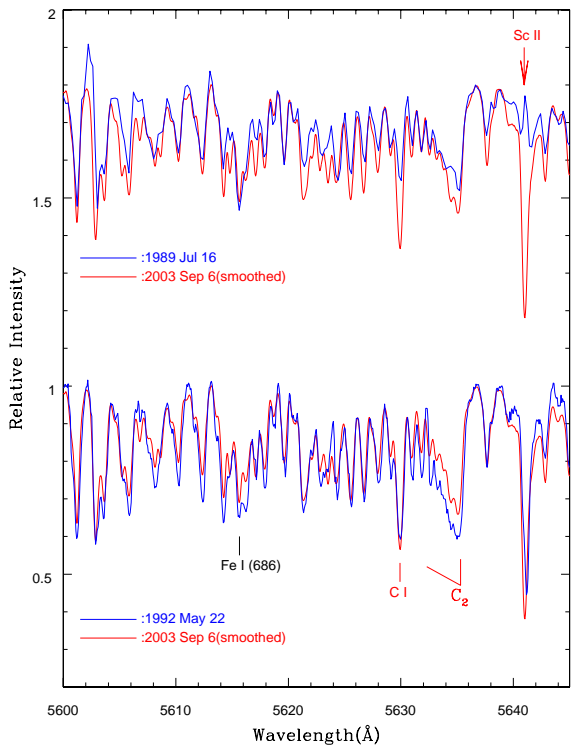}
\caption{The minimum light spectra of 1989 July 16 and 1996 May 22
 are compared
 with the
maximum light spectrum obtained on 2003 September 6 (red line) after smoothing
to the same  resolution. The sharp emission in  Sc\,{\sc ii} 
line in the 1989 spectrum is shown by the arrow. Most of the absorption lines 
are of lower depth in the 1989 spectrum relative to the
 maximum spectrum. The 1992 
spectrum obtained at a shallow minimum shows the C$_2$ band and low 
excitation lines to be enhanced in absorption but the sharp Sc\,{\sc ii}
 emission is
effectively absent.} 
\end{figure}

 The absorption lines are less strong than
at maximum light  Unlike R CrB where the Na\,{\sc i} sharp
emission line is very strong, the sharp emission at NaD is inconspicuous for
V CrA.
 The weak broad Na\,{\sc i} D emission extends symmetrically
 relative to the stellar velocity (Figure 11) extending to about
$\pm$100 km s$^{-1}$.
 There is an indication of shell absorption present at -120 km s$^{-1}$ for
the  Na\,{\sc i} D lines.

\begin{figure}
\epsfxsize=8truecm
\epsffile{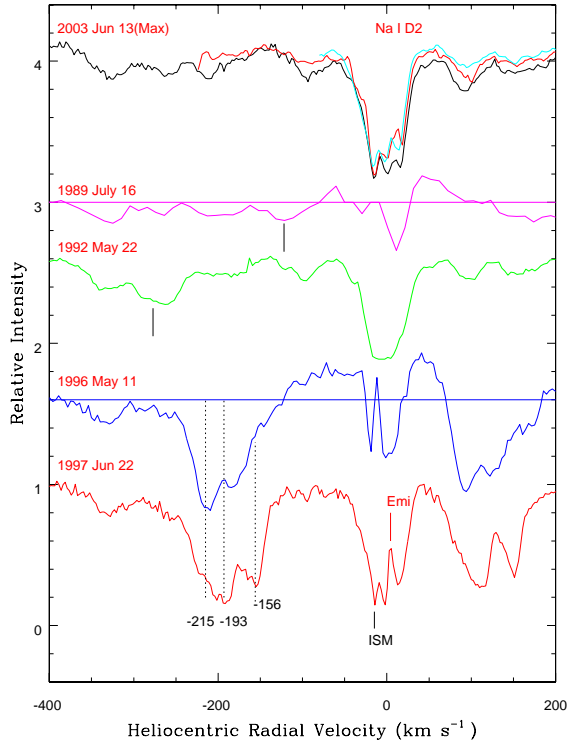}
\caption{The profiles of Na\,{\sc i} D2.  
The top spectrum shows
 the profiles on three occasions during
maximum light (2000 June 16 (blue), 2003 June 13 (red), 2003 September
6 (black)).
Slight variations in the stellar line are seen but, as expected,  the
interstellar (ISM) component
 $-14.7$ km s$^{-1}$ is not variable. The other four
spectra show the profile during various minima. Shell absorptions
during minima are marked. Sharp emission in the core
of the Na D line in the 1997 June 22 spectrum is identified.
The position of the continuum is shown for the 1989
July 16 (magenta line) and 1996 May 11 (red line) spectra
to highlight the broad emission.}
\end{figure}

  {\bf 1992 May 22}: When this spectrum was obtained, V CrA was less than
a magnitude below maximum light and the ensuing minimum was shallow
attaining a maximum depth of just 1.8 magnitudes.
 Not surprisingly perhaps, the spectrum is very similar to that
at maximum light.
 A comparison (Figure 10) of the 2003 September 6 maximum light spectrum (after 
 smoothing for the resolution difference) with the 1992 May 22 spectrum shows a 
 very good match  except for very low-excitation
absorption lines and C$_{\rm 2}$ Swan  bands  which are stronger.
Weak shell absorption in the Na\,{\sc i} D lines seems to be
 present at -270 km s$^{-1}$,  a remnant of ejected gas from earlier minima.

  {\bf 1996 May 11}: V CrA was well into a prolonged minimum that started about
 280 days before our observation and covered three declines and partial
 recoveries
  without reaching its usual maximum brightness (Figure 1).
  The spectrum was obtained at the begining of the first recovery phase.
  It mostly shows an absorption
 spectrum similar to that of  maximum light spectrum except for
 the presence of
  broad Na\,{\sc i} D emission lines and 
  strong shell absorption at $-215$ km s$^{-1}$ for
 lines of Na\,{\sc i} D (Figure 11) and K\,{\sc i}.
A comparison with R CrB during its 1995--1996 minimum is shown
 in Figure 12 for the Na\,{\sc i} D2 and the K\,{\sc i} 7698 \AA\ lines.
 The spectra of R CrB and V CrA were obtained about 220 and 280 days,
 respectively, after  the intial drop in light.
 Obviously, the very strong sharp Na\,{\sc i} D emission in R CrB is
 missing in the spectrum of V CrA even though the broad emission  
is of similar strength. Another interesting contrast is seen for
the K\,{\sc i}
line 7698 \AA. The line is a normal absorption line  at maximum  in
 R CrB and  the blueshifted shell absorption components, corresponding to 
 the strong Na\,{\sc i} D components, are not present.
The line in V CrA shows  
 blueshifted shell absorption  corresponding to the Na\,{\sc i} D
 lines.
A striking aspect of this and other spectra of V CrA in decline is that
the numerous sharp emission lines characteristic of R CrB (and other RCBs)
are largely absent from V CrA (Clayton et al 1992).
 
 \begin{figure}
\epsfxsize=8truecm
\epsffile{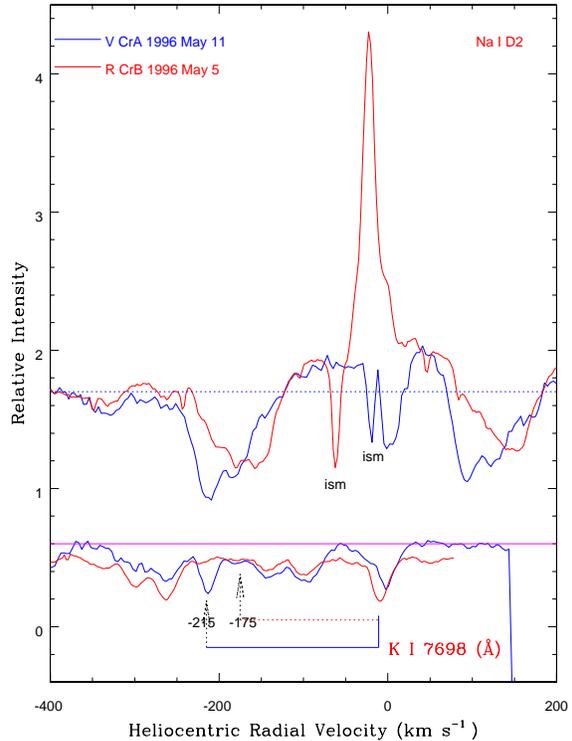}
\caption{Comparison of the profiles of the Na\,{\sc i} D2
 and K\,{\sc i} 7698 \AA lines in V CrA (blue) and R CrB (red) obtained
  when both stars were 4.8 magnitudes fainter than at maximum light and
 recovering from a deep minimum. Note the strong blueshifted shell
 absorption components to the Na\,{\sc i} D2 profiles with similar expansion
 velocities in both stars. 
The very strong sharp emission in the Na\,{\sc i} D2 line seen in
 R CrB is absent in V CrA.
 }
\end{figure}

  {\bf 1997 June 22}: The spectrum was obtained on the recovery phase
 of the second major drop in light in the prolonged minimum that started 
 around 1995 July. Subsequently, the star became fainter than V $\simeq$ 15.5
before recovering.
In most respects, the 1997 June 22 spectrum resembles a spectrum
taken at maximum light. Three features not attributable to a
stellar photosphere are present:  (i) absorption lines
of the C$_{\rm 2}$ Phillips system; (ii) the expected high velocity
 blue-shifted broad absorption components of the  Na\,{\sc i}  D
and the  K\,{\sc i} resonance lines; (iii) sharp emissions.

\subsection{Absorption lines of C$_{\rm 2}$ Phillips system}

Weak
 absorption lines of the C$_{\rm 2}$ Phillips system are present in the
spectrum of 1997 June 22 (Figure 13). This system arises from
the molecule's electronic ground state whereas the Swan system's
lower electronic state is a low-lying electronic state (Ballik \&
Ramsay 1963). In dilute gas, as found in a circumstellar shell,  the Phillips
system is the transition of choice for detecting C$_2$ molecules.
 We measured lines of the 3-0
 band in the  7760 \AA\ to 7880 \AA\ interval. This is the only spectrum in our
collection to cover the 3-0 band to show these absorption lines.

\begin{figure}
\epsfxsize=8truecm
\epsffile{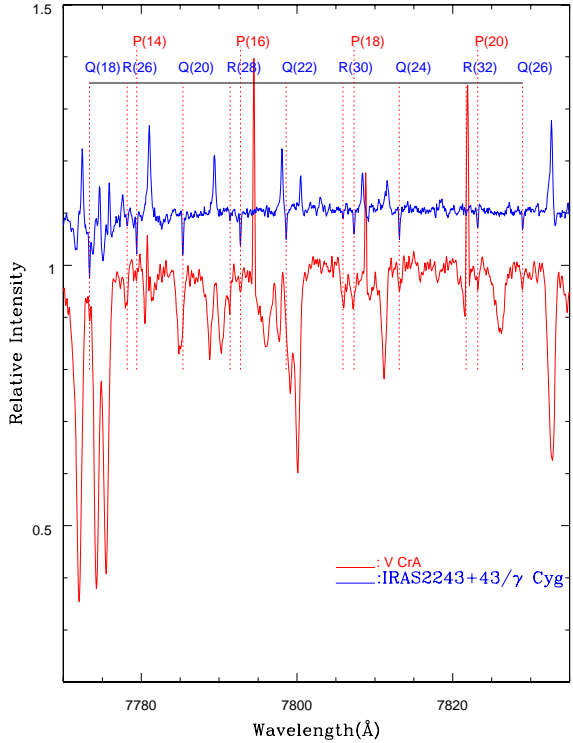}
\caption{The spectrum of V CrA obtained on 1997 June 22  (red) is shown
along with a spectrum of the  post-AGB star IRAS2243+43 (blue)
% (divided by $\gamma$ Cyg
%to remove the stellar lines and highlight the circumstellar lines)that
 show the absorption lines of Phillips 3-0 system identified by the.
red dotted lines.
 The 
emission-like features in the IRAS spectrum are  artefacts resulting from
the procedure used to ratio out the telluric lines.
The sharp emission lines in the spectrum of
V CrA are uncompensated terrestrial OH lines. }
\end{figure}

\begin{figure}
\epsfxsize=8truecm
\epsffile{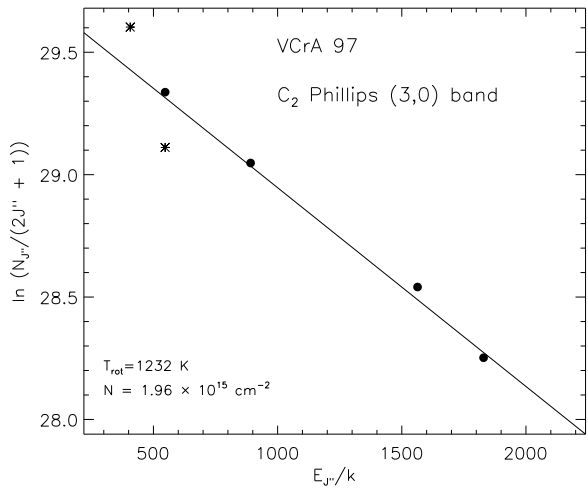}
\caption{Boltzmann plot of the absorption lines of the 3-0 band of the C$_2$
Phillips system from the spectrum of V CrA obtained on 1997 June 22.
 Star and dot symbols refer to P and Q branch lines respectively.}
\end{figure}

A rotational temperature of 1230$\pm$30 K is obtained (Figure 14)
using the molecular data given by Bakker et al. (1998).
The molecular  column
 density is 2 x 10$^{15}$ cm$^{-2}$.
 The radial velocity of the C$_2$ molecules is  $-$10.2 km s$^{-1}$ suggesting
 an expansion velocity of 4.6 km s$^{-1}$ relative to the velocity of the
main component
 (- 5.6 km s$^{-1}$) but only about 2 km s$^{-1}$ relative to the
star's systemic velocity.
Presence of cool C$_2$ molecules
suggests that the gas is at temperatures where dust
may condense.
 Similar detections of cool C$_2$ molecules
have been reported by us for V854 Cen (Rao \& Lambert 2000) and R CrB (Rao,
Lambert \& Shetrone 2006).

\subsection{Shell Absorption Components}

Blueshifted  absorption components to  the Na\,{\sc i} D lines and K\,{\sc i}
lines are common across the minimum light spectra, as noted above.
  The spectra of 1996 May 11 and 1997 June  22 are of particular interest as
they were obtained in sucessive minima during a prolonged minimum.
In Figures 10 and 11,
the stellar Na D2
lines appear near the systemic velocity of about $-8$ km s$^{-1}$.
The stellar line consists of two components on some spectra and is
blended with an ISM line of constant velocity and strength.
                               The blue-shifted shell
components are several in number with velocities between about
$-160$ km s$^{-1}$ to $-230$ km s$^{-1}$ with the strongest
absorption at $-190$ km$^{-1}$ . The component at $-156$ km s$^{-1}$
was not present in 1996 May 11  but emerged as an additional
component in 1997 June.
   The K\,{\sc i} 7698 \AA\ line clearly shows (Figure 15) 
the $-215$ km s$^{-1}$ component, which was strong in 1996 May 
and weakening  by 1997 June 22 and the $-193$ km s$^{-1}$ component that
was not present in 1996 May appearing with greater strength in 1997 June.

\begin{figure}
\epsfxsize=8truecm
\epsffile{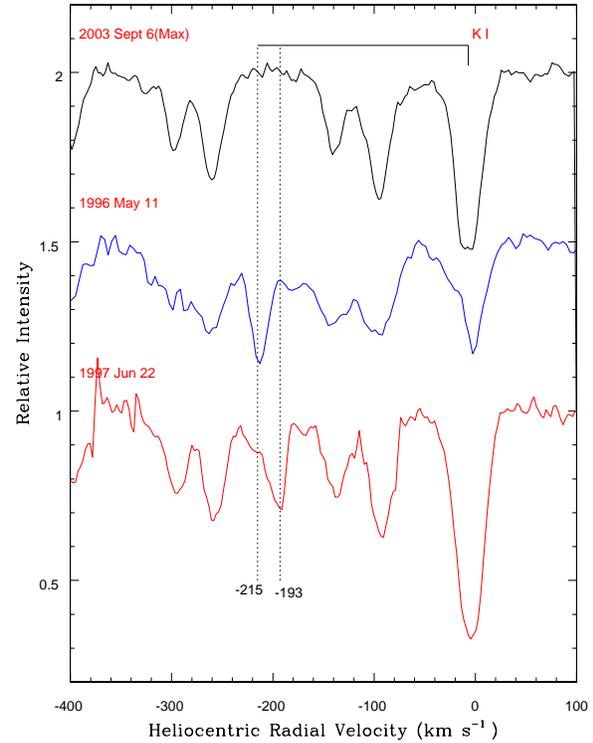}
\caption{The profiles of K\,{\sc i} 7698 \AA line and its blueshifted
shell absorption components at the 1996 and 1997 minima along with the 
maximum light spectrum. The shell components are absent at light maximum.
Note that the component at $-215$ km s$^{-1}$ which is strong at the
 1996 minimum
had weakened considerably by 1997 but a new shell absorption component at 
-193 km s$^{-1}$ emerged in 1997. The 1996 spectrum shows
the stellar K\,{\sc i} 7698 \AA line has been filled in by emission.} 
\end{figure}

\begin{figure}
\epsfxsize=8truecm
\epsffile{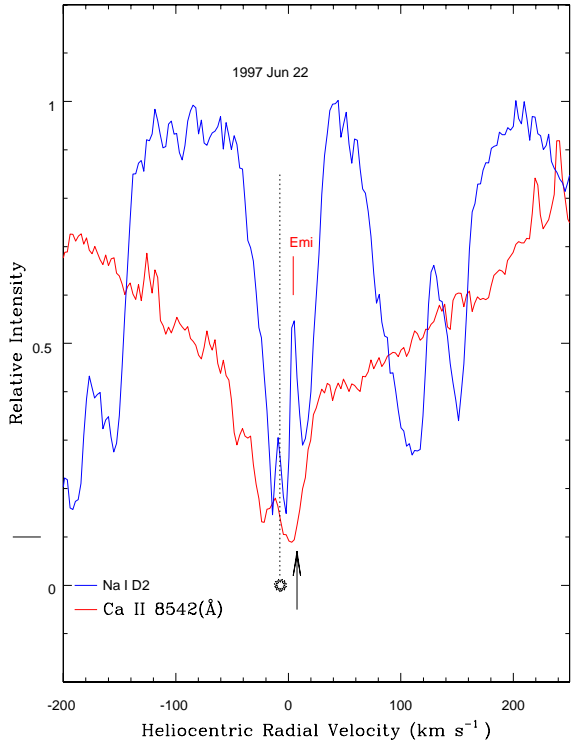}
\caption{The profiles of the  Na\,{\sc i}  D2  and Ca\,{\sc ii} 8542 \AA lines
from the spectrum obtained on 1997 June 22. The radial velocity scale is
heliocentric. Note the sharp emission present in Na\,{\sc i} D2 is absent 
from the Ca\,{\sc ii} line.  
} 
\end{figure}

 Rough estimates of the column densities of Na\,{\sc i} and K\,{\sc i} 
in the velocity range $-230$ to $-190$ km s$^{-1}$  
are made by using the doublet method 
(Spitzer 1968). These give the abundance ratio
$\log$(Na\,{\sc i}/K\,{\sc i}) $\simeq -0.4$ and $-0.6$ for the
two parts of the profile . The photospheric abundance analysis (Table 3)
gave the elemental abundance ratio as $-0.4$ which is fortuitously close to
the ratio from the shell lines.
The total column density of Na\,{\sc i} 
is 9.3 x 10$^{12}$ cm$^{-2}$. By assuming the Na to He ratio as estimated
earlier and the
size of the cloud based on its velocity of
expansion and its dispersion (from Na\,{\sc i} D lines), 
the cloud mass is estimated as 6.9 x 10$^{24}$ gm or
 3.4 x 10$^{-9}$ M$_{\odot}$.

\subsection{Sharp Emission lines}
 
   Even during this minimum  sharp emission lines, that are a characterisic 
of RCB stars at minimum,
 are only represented in the cores of Na\,{\sc i} D lines (Figure 16)
 in the spectral regions observed. The possibility exists that the sharp emission region was abscured by dust and only very strong emissions (eg.Na\,{\sc i} D)
 could be seen. In addition the level of excitation appears to be low in the
 sharp emission region such that only emission is present in Na\,{\sc i} D
 lines but not in Ca\,{\sc ii} triplet lines (Figure 15). The absorption lines
,particularly of low excitation ,seems to have redward components (Figure 3c).

\section{Discussion}

Important questions posed by RCB stars remain unanswered. 
The two most fundamental concern the origin of these H-deficient
stars, and the mechanisms by which a cloud of soot forms and
obscures the star. One supposes that the origins of the
stars will be betrayed by examination of the abundances of the
stellar atmospheres and that thorough spectroscopic and
photometric scrutiny of stars 
from maximum light into and through declines will provide the clues
to the understanding of cloud formation.

Two scenarios remain in play to account for a H-deficient luminous
star. In the first, a final He-shell flash in a post-AGB star on the
white dwarf cooling track creates a H-deficient luminous star 
that evolves rapidly to lower temperatures. This is dubbed the
`final flash' (FF) scenario. In the second, the H-deficient luminous
star is formed from the merger of a He and a C-O white dwarf. In the
close binary system with loss of energy by gravitational radiation, 
accretion of the He white dwarf  by the C-O white dwarf leads,
subject to restrictions such as the total mass should not exceed
the Chandrasekhar limit, to a H-poor supergiant with the C-O white
dwarf as its core. This is dubbed the `double-degenerate' (DD)
scenario.

 Evidence 
suggests that the DD rather than the FF scenario
provides the superior accounting for the elemental abundances of
C,N, and O for RCB stars and their likely relatives the extreme Helium 
(EHe) stars (Saio \& Jeffery 2002; Asplund et al. 2000; Pandey et al.
2006). Convincing support for the DD scenario as creators of the
cool hydrogen deficient (HdC) stars, another example of likely
relatives of RCBs,  is provided by Clayton et al.'s
(2005, 2007) discovery that $^{18}$O is very abundant in those
HdC's with strong CO vibration-rotation bands; $^{18}$O  may be
readily synthesized by $\alpha$-capture from abundant $^{14}$N
during the merger in the DD scenario but not made with ease in the
FF scenario. Several cool RCBs also show large amounts of $^{18}$O
(Clayton et al. 2007)  but warm RCBs such as V CrA and R CrB 
do not show CO bands in their infrared spectra (Tenenbaum et al. 2005).
The FF scenario does plausibly account for other stars. Most notable
among the FF candidates are FG Sge  and V4334 Sgr, also known
as Sakurai's object and V605 Aql (Clayton et al 2006). 
It is not impossible that RCBs from the
FF scenario lurk among the analysed sample attributed in the main to the
DD scenario.

\subsection{Composition of V CrA}

Our determination of V CrA's chemical composition fully confirms the
star's status as a minority RCB with the added important information that
V CrA is rich in $^{13}$C. Indeed, the estimated $^{12}$C/$^{13}$C
ratio is equal within the measurement uncertainties to the
ratio expected from running of the H-burning CN-cycle. This cycle
cannot have proceeded to equilibrium because the C/N ($\simeq 5$)
 ratio is much
greater than one instead of much less than one, as required of
equilibrium abundances of C and N. One supposes that C-rich gas
was mildly exposed to warm protons.

\begin{figure}
\epsfxsize=8truecm
\epsffile{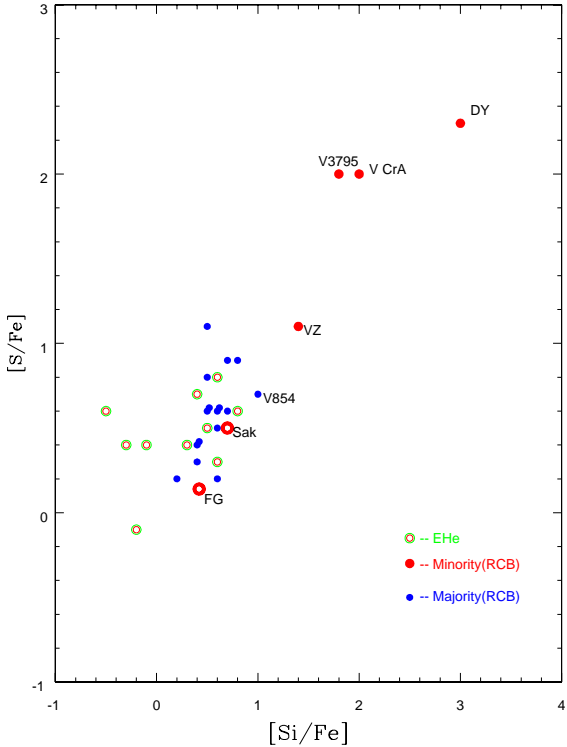}
\caption{The abundance ratio [Si/Fe] vs [S/Fe] for minority RCBs (red dots,
stars identified by name), majority RCBs (blue dots), EHe stars (yellow dots),
and the FF candidates FG Sge and Sakurai's object (unfilled red dots).
 }
\end{figure}

Pronounced
signatures of a minority RCB are the high Si/Fe and S/Fe ratios (Lambert
\& Rao 1994).
That these ratios are extraordinary for V CrA is shown in
Figure 17  where [Si/Fe] vs [S/Fe] is plotted for  V CrA (see Table 3),
R CrB and V2552 Oph (Rao \& Lambert 2003), other warm RCBs (Asplund
et al. 2000), V854 Cen (Asplund et al. 1998),
the hot RCB DY Cen (Jeffery \& Heber 1993),  Sakurai's object
(Asplund et al. 1997b), FG Sge (Gonzalez et al. 1998),
 and the extreme helium stars (EHEs)
(Pandey et al. 2001, 2006; Pandey \& Reddy 2006).
Three of the known minority RCBs -- V CrA, V3795 Sgr, and DY Cen --
 stand out in Figure 17 far
apart from the region occupied by the RCBs and the EHes.
It is noteworthy that
this trio span a wide range in effective temperature: V CrA at 6500 K,
V3795 Sgr at 8000 K, and DY Cen at 19500 K. Therefore, it seems
unlikely that the odd abundances are  artefacts resulting from a
misrepresentation of their atmospheres and systematic errors in the
abundance analyses.
 The fourth
known minority RCB is VZ Sgr which appears to form a bridge between the
first three and the other RCB and EHe stars that constitute the
majority of these H-poor stars. The mean [Si/Fe]$=+0.5$
 and [S/Fe]$=+0.6$ of the
majority RCBs are elevated by about $+0.2$ to $+0.3$ dex
 with respect to the values
found for unevolved disk stars of the sample's mean metallicity.
The magnitude of the elevation is possibly within the
range of systematic errors incurred by use of classical model
atmospheres and the assumption of LTE.

\begin{figure}
\epsfxsize=8truecm
\epsffile{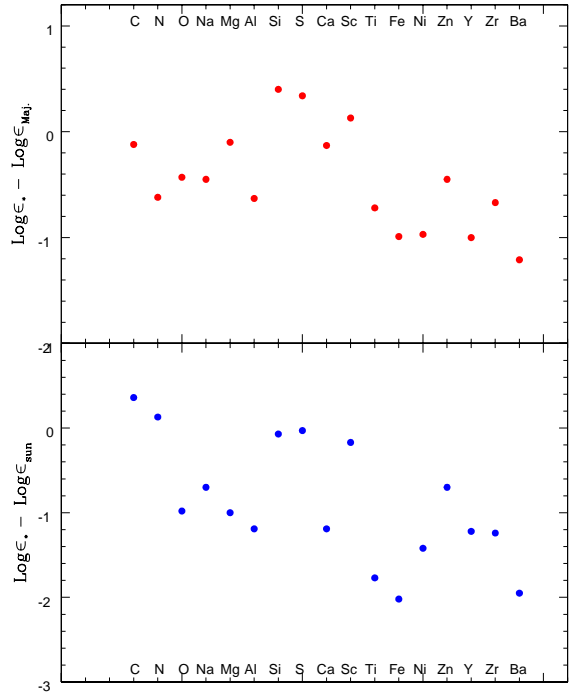}
\caption{The composition of V CrA. In the top panel, abundance differences
(in dex) are given for V CrA relative to the mean of the majority RCBs
from Asplund et al. (2000). In the bottom panel, abundance differences
(in dex) are given for V CrA relative to the solar system abundances from
Lodders (2003).
 }
\end{figure}

Two general comparisons of the composition of V CrA are offered
in Figure 18. In the top panel, the abundance differences are
presented between V CrA and the mean of the majority RCBs from
Asplund et al. (2000). This comparison may reduce the effect of the
carbon problem uncovered and discussed by Asplund et al. (2000).
The relative overabundance of Si, S, Sc and possibly Ca is
seen. These elements aside there is a general and roughly uniform
underabundance of elements in V CrA with respect to the mean abundances of the
majority RCBs.  In the lower panel, the comparison is between V CrA and the
solar system (Lodders 2003). Here, Si, S, and Sc stand out as
overabundant in V CrA.  As seen from Figure 18, the ratio of Si/Fe and
S/Fe far exceed the normal enhancement expected of a metal-poor
star. This exceptional excess does not apply to Mg, Ca, and Ti
which in normal  metal-poor stars share the small excess seen
by Si and S (say, [El/Fe] $\simeq +0.3$).
(C, N, and O in V CrA and all RCBs were altered in the course
of stellar evolution.) The other minority RCBs --
 V3795 Sgr, VZ Sgr, and DY Cen -- resemble V CrA in composition.
 V3795 Sgr is a close twin of V CrA but for a lower H
abundance by  at least four orders of magnitude.\footnote{The Ni
abundance in Table 3 is 0.7 dex lower than that given by Asplund et al.
With this abundance, V3795 Sgr has a Ni abundance that is 0.9 dex higher than
that of V CrA.} 
VZ Sgr is very similar to V CrA but for lower  [Si/Fe] and [S/Fe]
and a clear enrichment of the $s$-process elements Y, Zr, and Ba.
   
An intriguing clue to the origin of minority RCBs may be the
demonstration here that V CrA is rich in $^{13}$C. In contrast,
the warm majority RCBs probably, the cool RCBs certainly, and the HdC stars
also certainly are
not $^{13}$C-rich;
see, for example, Cottrell \& Lambert (1982b), 
 Warner (1967),  Clayton (1996), Tenenbaum et al. (2005),
and Clayton et al. (2007).
Most unfortunately, 
the  minority RCBs V3795 Sgr and DY Cen
are too hot to provide detectable C$_2$ bands and, hence, it is
unknown whether
their atmospheres are as rich in $^{13}$C as is V CrA's.
It might be supposed that synthesis of $^{13}$C in the case of V CrA
may be related to survival of hydrogen in its atmosphere,
the H abundance, although depressed by four orders of magnitude from its
assured initial value, is about two orders of magnitude greater than in
other warm RCB stars. Yet, V854 Cen with a factor of ten more hydrogen is not 
$^{13}$C-rich: $^{12}$C/$^{13}$C $\geq 30$ is the limit we set from the 1-0
Swan band.

With the separation of the minority stars from the majority, it is
difficult not to suspect that the minority followed an
evolutionary path with a branch or major diversion not open to the
majority. The similarity in abundances among the
minority stars suggests that, whatever the path, it led
to definitive abundances for Si, S, and Sc for the traveller.
 Events along the minority's path  may have been a
mixture of nuclear and chemical effects. The low  $^{12}$C/$^{13}$C
ratio is an example of a nuclear effect perhaps specific to the
minority stars; majority RCBs showing strong Swan bands do not
 show a detectable 4747 \AA\ $^{12}$C$^{13}$C bandhead.
In contrast to the nuclear process betrayed  by the high $^{13}$C
abundance, the run of abundance anomalies is not so obviously
attributable to nucleosynthesis. The high (solar) abundances of
Si, S, and Sc must be reconciled with lower
abundances of  Na, Mg, Al, Ca, Ni, Zn, Y and Zr, and even   
lower  abundances of Ti, Fe, and Ba. This abundance
pattern is shared  with the other minority
stars. One speculation is that the minority RCBs result from a variant
of the DD scenario - for example, a He white dwarf merger with a ONeMg
white dwarf, the latter resulting from an intermediate mass AGB star.
Nucleosynthesis in the merger may be capable of producing Si and S
from the mixing of He with the surface of the ONeMg white dwarf.
An intriguing observation of possible relevance is that  
analyses of the ejecta of ONeMg novae have shown several
examples with Si enhancements and one with a S enhancement; sulphur lines are 
generally not detected. For V838 Her, Schwarz et al. (2007) reported Si
and S enriched by factors of 5 and 22, respectively, with no
enhancement of Mg. Elements heavier than S were not reported but Al
was found to be overabundant by a factor of 18, an enrichment counter
to the results for the minority RCBs.  

The pattern is not obviously
attributable to a chemical effect such as is the case with
the dust-gas separation
process that creates abundance anomalies seen in
some RV Tauri variables (Giridhar et al. 2005).  Affected RV Tauri
with an oxygen-rich atmosphere  provide an abundance pattern that is not
that shown by the minority RCBs. The minority stars presently
have a carbon-rich atmosphere. The difference between O-rich and C-rich
gas  results in
grains of different composition (e.g., silicate vs graphite) forming
from cool gas and, hence, different compositions for the
remaining gas.
% It is far from obvious that this difference
%can account for the minority's abundance anomalies. 
If the dust-gas separation
were to operate in gas from the RCB star, such that
 SiC grains should be abundant and their
expulsion from gas would result in a Si {\it underabundance} when gas
was accreted by the star. The ISO spectra of the three RCB stars observed
R CrB, RY Sgr and V854 Cen do not indicate any features attributable
to SiC grains (Lambert et al 2001). The puzzle of 
the minority RCBs' abundance anomalies remains.

\subsection{The sooty outer atmosphere}
     
        A stellar wind, as sensed by the He\,{\sc i} 10830 \AA\ and
O\,{\sc i} 7771 \AA profiles,  is fed by the atmosphere.
  In addition, pulsations 
 are present during light maximum with mild shocks 
 propagating through the atmosphere as evidenced by the line doubling.
The shocks may feed the stellar wind.
 Probably, when a strong shock occurs the  post-shocked region
 might reach temperature sufficiently cool 
 for the dust to condense (Woitke et al. 1996). 
 How the dust grains grow  and clouds form is not quite clear;
Woitke (2006) and Woitke \& Niccolini (2004) have modelled dust formation
in carbon-rich AGB stars. Radiation pressure on dust grains and dust-gas
collisional coupling lead to ejection of clouds that are seen by the
shell blue-shifted absorption components of the Na\,{\sc i} D and other
lines.
 These components seen
in spectra taken at minimum and during recovery
to maximum are  a common feature of majority RCBs and apparently too
of minority RCBs (Rao et al. 1999), The absorbing gas would
appear to be associated with the dust causing the ongoing
decline.

Presence of C$_2$ molecules at low rotational temperatures may also
be a common and not unexpected feature of RCBs in decline. In addition
to their detection here in V CrA, the
molecules have been detected for V854 Cen (Rao \& Lambert 2000)
and R CrB (Rao, Lambert, \& Shetrone 2006). In all cases, the
velocity of the C$_2$ molecules is low relative to the systemic velocity.
Molecules have not been detected accompanying the high-velocity Na
atoms.

Spectra described here show differences in the emission line spectrum at 
minimum light between V CrA and well studied majority RCB stars,
particularly R CrB and RY Sgr. Especially noticeable is the weakness
of the sharp emission lines (relative to the stellar continuum) from
V CrA, as highlighted by the absence of a sharp emission component to the
Na D lines. Since the location of the emitting region is uncertain for
even the well studied RCBs, it is difficult to interpret this seemingly
unusual aspect of V CrA. The broad lines of which the Na D lines are the sole
example seen in V CrA appear to be of similar strength (relative to the stellar
continuum) in V CrA and R CrB (Figure 11).  

The stellar absorption line spectrum seen when the star has faded
by  three to five magnitudes is a replica of the spectrum at maximum light;
there is no indication that the spectrum is `veiled', i.e., the absorption
lines are much shallower and broader and even slightly Doppler shifted. 
Since such
veiling was seen in the 1995--1996 minimum of R CrB only when the star was
7 or more magnitudes below maximum light, the absence of veiling for V CrA
observed no more than five magnitudes from maximum may be not be a
surprise. Although some clouds may be optically thick, the star may have
been seen through optically thin clouds or gaps between clouds. 

\section{Concluding remarks}

Differences  between the spectra of RCBs at minimum light
encourage us to
continue our  monitoring of V CrA and other RCBs. Studies of the initial
stages of a decline should reveal clues to the trigger that sets off a
decline. In this era when photometric observations by amateur astronomers
are reported on the internet almost instantaneously, spectroscopic and
other follow-up observations are limited by access to suitable 
telescopes. The advent of queue scheduling is smoothing the path to
obtaining the follow-up observations. Development of a consortium of
observers would also ease the situation. Observations of the RCBs in the
deepest of minima are likely to provide novel data on their
extended atmospheres, especially on the enigmatic broad lines which, the
Na D lines  apart, have been studied in detail with high-quality spectra
only in the case of R CrB. For RCBs, aside from the three brightest
stars R CrB, RY Sgr, and V854 Cen, a large telescope will be needed to
acquire quality spectra at the faint magnitudes expected to reveal the
broad lines.

\section{Acknowledgements}
                                                                                     
We acknowledge with thanks the variable star observations from the 
AAVSO  database.
This research has made use of the SIMBAD database, operated
at CDS, Strasbourg, France. Our sincere thanks to Suchitra Balachandran,
Jennifer Simmerer, David Yong, Eswar Reddy for securing  observations
at our request. We would also like to thank Martin Asplund for supplying
line lists, programmes, and instructions on how
to compute the  C$_2$ bands. Thanks are  due to David Yong for
his  assistance. This research has been supported in part by a grant from
the Robert A. Welch Foundation of Houston, Texas. We would like to thank
the refree, Geoff Clayton, for useful comments.

\end{document}